\documentclass[10pt,journal]{IEEEtran}
\usepackage{algorithmic}
\usepackage{algorithm}
\usepackage{array}
\usepackage[caption=false,font=normalsize,labelfont=sf,textfont=sf]{subfig}
\usepackage{textcomp}
\usepackage{stfloats}
\usepackage{url}
\usepackage{verbatim}
\usepackage{graphicx}
\usepackage{cite}
\usepackage{verbatim}

\newcommand{\getsr}{\ensuremath{\leftarrow_\$}}

\newcommand{\Zq}{\mathbb{Z}_q}

\newcommand{\bits}[1]{\{0,1\}^{#1}}
\newcommand{\secpar}{\lambda}
\mathchardef\mhyphen="2D

\newtheorem{theorem}{\textbf{Theorem}}

\newtheorem{remark}{\textbf{Remark}}

\newcommand{\AuC}{\mathrm{AuC}}
\newcommand{\UE}{\mathrm{UE}}
\newcommand{\HgNB}{\mathrm{HgNB}}
\newcommand{\gNB}{\mathrm{gNB}}

\newcommand{\numParties}{\ensuremath{n_P}}
\newcommand{\numSessions}{\ensuremath{n_S}}

\newcommand{\Alg}[1]{\ensuremath{\mathsf{#1}}}
\newcommand{\SanSig}{\Alg{SanSig}}
\newcommand{\AuthEnc}{\Alg{AE}}

\newcommand{\KeyGen}{\Alg{KGen}}
\newcommand{\Sign}{\Alg{Sign}}
\newcommand{\Sanit}{\Alg{Sanit}}
\newcommand{\Verify}{\Alg{Verify}}
\newcommand{\proof}{\Alg{Proof}}

\newcommand{\Judge}{\Alg{Judge}}

\newcommand{\Enc}{\Alg{Enc}}
\newcommand{\Dec}{\Alg{Dec}}
\newcommand{\KDF}{\Alg{KDF}}
\newcommand{\DDH}{\Alg{DDH}}
\newcommand{\KGenacc}{\Alg{KGen_{acc}}}
\newcommand{\Genacc}{\Alg{Gen_{acc}}}
\newcommand{\Updateacc}{\Alg{Update_{acc}}}
\newcommand{\Nwitcreate}{\Alg{NonWitCreate}}
\newcommand{\Nwitupdate}{\Alg{NonWitUpdate}}
\newcommand{\Verifyacc}{\Alg{Verify_{acc}}}

\newcommand{\challenger}{\mathcal{C}}
\newcommand{\adversary}{\mathcal{A}}
\newcommand{\chal}{\ensuremath{\mathcal{C}}}

\newcommand{\adv}{\ensuremath{\mathcal{A}}}

\newcommand{\Unlink}{\Alg{Unlink}}
\newcommand{\Prot}{\Pi}

\newcommand{\testsess}{\pi_i^s}
\newcommand{\session}{\pi}
\newcommand{\cleanpredicate}{\mathbf{clean}}

\newcommand{\MA}{\ensuremath{\mathsf{MA}}}
\newcommand{\KIND}{\ensuremath{\mathsf{KIND}}}

\newcommand{\pk}{\ensuremath{\mathit{pk}}}
\newcommand{\sk}{\ensuremath{\mathit{sk}}}
\newcommand{\adm}{\ensuremath{\mathit{ADM}}}
\newcommand{\mod}{\ensuremath{\mathit{MOD}}}
\newcommand{\cert}{\ensuremath{\mathit{CERT}}}
\newcommand{\pid}{\ensuremath{\mathit{PID}}}

\newcommand{\skacc}{\ensuremath{\mathit{sk_{acc}}}}
\newcommand{\EID}{\mathrm{EID}}
\newcommand{\HID}{\mathrm{HID}}
\newcommand{\RID}{\mathrm{RID}}
\newcommand{\PID}{\mathrm{PID}}
\newcommand{\TID}{\mathrm{TID}}
\newcommand{\ZUID}{\mathrm{ZUID}}
\newcommand{\RUID}{\mathrm{RU\mhyphen ID}}
\newcommand{\RL}{\mathit{RL}}

\newcommand{\Corrupt}{\ensuremath{\Alg{Corrupt}}}
\newcommand{\StateReveal}{\Alg{StateReveal}}

\newcommand{\Test}{\Alg{Test}}

\newcommand{\Adv}[2]{\ensuremath{\mathbf{Adv}^{#1}_{#2}}}
\newcommand{\Exp}[2]{\ensuremath{\mathbf{Exp}^{#1}_{#2}}}

\newcommand{\ClientAction}[1]{ 
	\node[right] at (\InitX, \Y) {#1};
}
\newcommand{\ServerAction}[1]{
	\node[left] at (\RespX, \Y) {#1};
}

\newcommand{\AdversaryAction}[1]{
	\node at ($1/2*(\InitX, \Y)+1/2*(\RespX, \Y)$) {#1};
}
\newcommand{\ClientToServer}[3][->]{
	\NextLine[0.5]
	\draw[#1] (\ArrowLeft,\Y) -- node[above] {#2} node[below] {#3} (\ArrowRight,\Y) ;
	\NextLine[0.5]
}
\newcommand{\ServerToClient}[3][->]{
	\NextLine[0.5]
	\draw[#1] (\ArrowRight,\Y) -- node[above] {#2} node[below] {#3} (\ArrowLeft,\Y) ;
	\NextLine[0.5]
}
\newcommand{\ClientToAdversary}[3][->]{
	\NextLine[0.5]
	\draw[#1] (\ArrowLeft,\Y) -- node[above] {#2} node[below] {#3} (\ArrowCenter,\Y) ;
	\NextLine[0.5]
}
\newcommand{\ServerToAdversary}[3][->]{
	\NextLine[0.5]
	\draw[#1] (\ArrowRight,\Y) -- node[above] {#2} node[below] {#3} (\ArrowCenter,\Y) ;
	\NextLine[0.5]
}

\newcommand{\AdversaryToClient}[3][->]{
	\NextLine[0.5]
	\draw[#1] (\ArrowCenter,\Y) -- node[above] {#2} node[below] {#3} (\ArrowLeft,\Y) ;
	\NextLine[0.5]
}
\newcommand{\AdversaryToServer}[3][->]{
	\NextLine[0.5]
	\draw[#1] (\ArrowCenter,\Y) -- node[above] {#2} node[below] {#3} (\ArrowRight,\Y) ;
	\NextLine[0.5]
}

\newcommand{\NextLine}[1][1.0]{
	\pgfmathparse{\Y+#1}
	\edef\Y{\pgfmathresult}
}
\newcommand{\linkgame}[2]{\hyperref[#1]{G#2}}

\newcounter{Bdversary}

\newcommand\orcidicon[1]{\href{https://orcid.org/#1}{\mbox{\scalerel*{
\begin{tikzpicture}[yscale=-1,transform shape]
\pic{orcidlogo};
\end{tikzpicture}
}{|}}}}

\newcommand*{\addFileDependency}[1]{
  \typeout{(#1)}

  \IfFileExists{#1}{}{\typeout{No file #1.}}
}

\newcommand*{\myexternaldocument}[1]{%
    \externaldocument{#1}%
    \addFileDependency{#1.tex}%
    \addFileDependency{#1.aux}%
}

\usepackage{multirow}
\usepackage{float}
\usepackage{adjustbox}
\usepackage{tikz}
\usetikzlibrary{svg.path}
\usetikzlibrary{calc}
\usepackage{scalerel}
\usepackage{color}
\usepackage{soul,xcolor}
\usepackage{makecell}
\newif\iffullversion
\newif\ifsubmissionversion
\fullversiontrue
\submissionversionfalse
\usepackage{academicons}
\usepackage{bm} 
\usepackage{amsfonts}
\usepackage{enumitem}
\usepackage{subfiles} 
\definecolor{orcidlogocol}{HTML}{A6CE39}
\usepackage{xr}
\usepackage{xcite}

\definecolor{orcidlogocol}{HTML}{A6CE39}
\tikzset{
  orcidlogo/.pic={
    \fill[orcidlogocol] svg{M256,128c0,70.7-57.3,128-128,128C57.3,256,0,198.7,0,128C0,57.3,57.3,0,128,0C198.7,0,256,57.3,256,128z};
    \fill[white] svg{M86.3,186.2H70.9V79.1h15.4v48.4V186.2z}
                 svg{M108.9,79.1h41.6c39.6,0,57,28.3,57,53.6c0,27.5-21.5,53.6-56.8,53.6h-41.8V79.1z M124.3,172.4h24.5c34.9,0,42.9-26.5,42.9-39.7c0-21.5-13.7-39.7-43.7-39.7h-23.7V172.4z}
                 svg{M88.7,56.8c0,5.5-4.5,10.1-10.1,10.1c-5.6,0-10.1-4.6-10.1-10.1c0-5.6,4.5-10.1,10.1-10.1C84.2,46.7,88.7,51.3,88.7,56.8z};
  }
}

\usepackage[hidelinks]{hyperref}

\myexternaldocument{appendices}

\begin{document}

\title{Privacy-aware Secure Region-based Handover \\for Small Cell Networks in 5G-enabled \\ Mobile Communication}

\author{Rabiah Alnashwan$^{\orcidicon{0000-0001-9156-1679}}$, Prosanta Gope$^{\orcidicon{10000-0003-2786-0273}}$ \textit{Senior Member, IEEE} , Benjamin Dowling$^{\orcidicon{0000-0003-3234-6527}}$
\IEEEcompsocitemizethanks{
\IEEEcompsocthanksitem R. Alnashwan is with the Department of Computer Science, University of Sheffield and the Department of Computer Science, Imam Mhammad In Saud Uiversity, Riyadh, Saudi Arabia. (E-mail: RAlnashwan@imamu.edu.sa/RAlnashwan1@sheffield.ac.uk)
\IEEEcompsocthanksitem P. Gope, and B. Dowling  are with Department of Computer Science, University of Sheffield, Regent Court, Sheffield S1 4DP, United Kingdom.
(E-mail: prosanta.nitdgp@gmail.com/p.gope@sheffield.ac.uk; b.dowling@sheffield.ac.uk)

}
}

\markboth{A\MakeLowercase{lnashwan et al.}\ }%
{Shell \MakeLowercase{\textit{Alnashwan et al.}}: A Sample Article Using IEEEtran.cls for IEEE Journals}

\maketitle

\begin{abstract}

The 5G mobile communication network provides seamless communications between users and service providers and promises to achieve several stringent requirements, such as seamless mobility and massive connectivity. Although 5G can offer numerous benefits, security and privacy issues still need to be addressed. For example, the inclusion of small cell networks (SCN) into 5G brings the network closer to the connected users, providing a better quality of services (QoS), resulting in a significant increase in the number of Handover procedures (HO), which will affect the security, latency and efficiency of the network. It is then crucial to design a scheme that supports seamless handovers through secure authentication to avoid the consequences of SCN. To address this issue, this article proposes a secure region-based handover scheme with user anonymity and an efficient revocation mechanism that supports seamless connectivity for SCNs in 5G. In this context, we introduce three privacy-preserving authentication protocols, i.e.,  initial authentication protocol, intra-region handover protocol and inter-region handover protocol, for dealing with three communication scenarios. To the best of our knowledge, this is the \emph{first} paper to consider the privacy and security in both the intra-region and inter-region handover scenarios in 5G communication.  Detailed security and performance analysis of our proposed scheme is presented to show that it is resilient against many security threats, is cost-effective in computation and provides an efficient solution for the 5G enabled mobile communication.

\end{abstract}


\section{Introduction}
\IEEEPARstart{T}{he} number of connected devices is continuously increasing, and forecasts predict to reach up to 50 billion connected devices worldwide in 2030 \cite{holst_number_2021}. With this continuous growth, cellular networks have significantly evolved over the generations (1G-5G). The recently deployed 5G network introduces stringent requirements to cope with the increased capacity of connected devices and provide enhanced QoS. These requirements include reduced latency and costs, lower energy consumption, increased network capacity, high data rates, and scalable device connectivity. To achieve all these requirements, 5G mobile communication networks provide a unified, programmable software‐centric infrastructure by merging recently developed network technologies, such as cloud computing, software-defined networking (SDN), network function virtualisation (NFV) and Ultra-dense Small Cell Networks (SCN) \cite{cao_survey_2020}.

Ultra-dense SCN technology achieves several 5G requirements: high network density, capacity, and spectrum efficiency. Ultra-dense SCN here refers to low-powered cellular radio access nodes. SCNs aim to increase the density of wireless cells/nodes and reduce the coverage area to approximately 10-100 metres. The increased density of nodes increases network capacity and brings the network closer to the connected users, providing better network throughput with low-powered transmission and increasing the spectrum efficiency. However, to access the new advantages of SCNs, users will need to hop between cells frequently (as the individual node coverage is significantly smaller), which is a process referred to as ``Handovers'' (HO). 
Transferring a device's ongoing wireless connection from one cell operated by a base station (BS) to another cell/ BS is commonly known as a handover process in mobile wireless communication. 

Although the usage of SCNs in 5G is advantageous in terms of utilising network resources and bringing the user closer to the network, it introduces more latency, security and privacy issues to the network, caused by the increased frequency of HO events in 5G. Since the 3GPP group \cite{3rd_generation_partnership_project_3gpp_security_2020} did not address the possible impact of SCNs on 5G-AKA and HO protocols, this leads to new protocols being introduced to reduce latency caused by frequent HOs in SCNs  \cite{fan_rehand_2020,zhang_robust_2021,yan_lightweight_2021}. 
In any cellular network (such as 5G), a secure authentication protocol plays a major role in ensuring users' legitimacy and also establishment of secure symmetric keys between users and services, enabling secure communication after that point. Generally, authenticating communication networks is carried out via Authentication and Key Agreement (AKA) protocols. For 5G authentication, the 3rd Generation Partnership Project GPP(3GPP) specifies two main AKA protocols: 5G-AKA  \cite{3rd_generation_partnership_project_3gpp_security_2020} and Extensible Authentication Protocol (EAP-AKA') \cite{arkko_improved_2009}. Both protocols are relatively similar, with some minor differences in the message flow, and so we focus on 5G-AKA in this work. The 5G-AKA protocol is executed between three main participants, a User (Equipment) UE, a Serving Network SN and the Home Network (HN), and introduces new network algorithms and protocols to handle the authentication and handover procedures between these parties.  The HN is responsible for issuing and maintaining users' unique information such as their telephone number, long term ID (known as their SUPI), long term shared symmetric key, and other UE-related information. Usually, UEs can access services provided by their HN through an intermediate semi-trusted party known as a serving network (SN). For UEs to authenticate to HNs, first \emph{user registration} must occur, where long-term symmetric keys and user credentials (SUPIs and SUCIs) are established between both parties. Now, UEs can securely \emph{authenticate} to HNs and vice-versa by using the 5G-AKA protocol \cite{basin_formal_2018}. In 5G-AKA, UEs first submit their (encrypted) SUCI to the HN, allowing the HN to recover the SUPI and identify the UE. The HN then sends a challenge to the UE, followed by a response from the UE. Thus, as seen by the SUCI construction, privacy is considered an equally vital security property in the 5G-AKA authentication and HO protocols. These protocols attempt to preserve user anonymity and untraceability, (often referred to as \emph{strong anonymity}), which are essential for preventing both an unauthorised disclosure of user identity as well as a linking of network activity to a single party. A considerable amount of research has focused on enhancing and analysing 5G-AKA \cite{cremers_component-based_2019, braeken_symmetric_2020, basin_formal_2018, borgaonkar_new_2018}. These works have addressed the security, privacy and efficiency in 5G-AKA authentication and HO protocols, identifying several weaknesses in 5G-AKA, such as identity replay attacks, attacks on sequence number confidentiality (breaking untraceability) and confusion attacks. 
 
 On the other hand, handover is also vital in mobile communication networks, especially in the SCN scenario. The insurance of a continual network connection with the massive deployment of BSs/cells arouses serious challenges to handover procedures in 5G. Therefore, 3GPP specifies a number of handover protocols that address the most common handover scenarios in 5G \cite{3rd_generation_partnership_project_3gpp_security_2020}. From these protocols, we classify them into two main categories: 1) homogeneous handovers and 2) heterogeneous handovers. Homogeneous handovers are operated between BSs within 5G networks, whereas the heterogeneous handovers are done between different networks,i.e. 5G, WiFi and LTE. In this work, we only consider homogeneous handovers within the 5G network. We can further divide the 5G  homogeneous handover into sub category based on the used interface defined by the 3GPP standard: Xn-based and N2-based handover. In Xn-based handover, requests are sent directly between two BSs over a pre-defined Xn interface. On the contrary, the N2 handover does not have direct communication between BSs. Handover requests instead are sent to the Access and Mobility Management Function (AMF), which resides in the core network, over the N2 interface. Peltonen, Sasse and Basin \cite{peltonen_comprehensive_2021} present a comprehensive security analysis of these two handover protocols specified in the 5G standard. Notwithstanding the 5G handover protocols, our proposed scheme aims and addresses the inclusion of SCN in 5G networks and introduces a scheme that can provide higher security while preserving users privacy in a region-based framework.
 
 There have been several handover authentication schemes proposed in the literature to overcome some of the vulnerabilities in 5G and also enhance the performance of authentication and HO protocols \cite{fan_rehand_2020,zhang_robust_2021,yan_lightweight_2021}. However, to the best of our knowledge, there is no such HO protocol that can achieve  \emph{strong anonymity} with \emph{perfect forward secrecy} and \emph{secure inter-region handover} for HO authentication in 5G. Therefore, there is a need for authentication and HO protocol that achieves the desired security properties mandated in 5G networks, such as mutual authentication, user anonymity, user untraceability, extending to SCN settings and achieving seamless user mobility \emph{in} and \emph{between} regions.

\subsection{Related Works}
A considerable amount of existing literature analyses 5G-AKA from different security and privacy perspectives \cite{cremers_component-based_2019,braeken_symmetric_2020,basin_formal_2018}. These studies have identified several weaknesses in the current version of 5G-AKA. Basin et al. \cite{basin_formal_2018} provide a comprehensive analysis of the 5G-AKA and discovered some underspecified security requirements in the original specifications.  This work identified traceability attacks against the 5G-AKA under the active adversaries. Cremers and Dehnel-Wild \cite{cremers_component-based_2019} analysed 5G-AKA from the security perspective, which revealed a confusion attack. This attack takes advantage of the identity misbinding property to launch an impersonation attack. Borgaonkar et al. \cite{borgaonkar_new_2018} also found a logical vulnerability, which breaks the confidentiality of the sequence number due to the usage of XOR and lack of randomness. Braeken \cite{braeken_symmetric_2020}, on the other hand analysed the 5G-AKA from the privacy perspective. The author discovered that a simple identity replay attack presented against several AKA protocols \cite{fouque_achieving_2016} also threatens 5G-AKA.  Accordingly, this work proposed an efficient 5G-AKA protocol that overcomes this attack, as well as addressing location privacy and linkability attacks. However, the proposed protocol provides only in-session unlinkability but not full unlinkability (between sessions): unlinkability holds only if all authentication attempts are successful, otherwise actions can be linked due to the un-updated GUTI values \cite{3rd_generation_partnership_project_3gpp_security_2020}. In addition, the same user location breach is not solved in the proposed protocol. Recently, Zhang et al. \cite{zhang_robust_2021}, Fan et al.\cite{fan_rehand_2020}, and Yan and Ma\cite{yan_lightweight_2021} proposed novel HO and authentication protocols specifically for the 5G environment. 
In 2020, Fan et al. \cite{fan_rehand_2020} proposed a secure region-based handover scheme (ReHand) with user anonymity and fast revocation for SCNs. This protocol supports a fast authentication only for users roaming inside a region only, i.e., between HgNBs belonging to the same gNB, using a  group secret key. In the initial authentication phase, which will be triggered during inter-region HO, users will send their pseudo-IDs to the AuC, update it and sending it back to the user. However, the ReHand protocol is susceptible to undetected desynchronization in the updated pseudo-ID. ReHand also provides a membership revocation mechanism that is realized by Nyberg’s one-way accumulator \cite{nyberg_fast_1996}. However, Nyberg’s accumulator efficiency is lower than other accumulators, and it is not dynamic, meaning AuC has to regenerate the revocation list and send it back to regions whenever a user is added or removed from that list, which negatively impacts the protocol efficiency. Next, in 2021 Zhang et al. \cite{zhang_robust_2021} proposed a universal HO and authentication scheme (RUSH) that exploits chameleon hash functions, blockchain, and Elliptic-Curve Diffie-Hellman (ECDH) key exchange. In this protocol, users are registered with the network using their actual identities and CH values. Then the authentication server stores users' identities locally and CH values in the Blockchain. Although the protocol supports universality (heterogeneous networks) realized by the Blockchain, user revocation and reply attacks were not considered in RUSH protocol. Additionally, some aspects of
blockchain in terms of security and performance have been
overlooked. Finally, recently Yan and Ma\cite{yan_lightweight_2021} proposed an efficient handover authentication protocol (LSHA) based on neighbouring BSs in the 5G network. In LSHA, each base station (gNB) has a secret key and session key with neighbouring gNBs generated by the AMF, which are used to secure the handover procedures. Although LSHA protocol is secure against DoS attacks and de-synchronization attacks, it only supports partial PFS and strong anonymity in the HO due to the dependence on the 5G-AKA specified by the 3GPP standard.

\subsection{Motivation and Contributions}
Despite all previous works in this domain, none of the existing protocols in the literature have considered all the 5G's requirements of a fast, secure, privacy-preserving and reliable HO authentication scheme. Most importantly, none of the existing 5G-AKA schemes have considered a secure inter-region HO scenario in the 5G networks. Therefore, there is a need for an authentication and HO scheme that achieves the desired security properties in 5G networks (explained in \ref{subsec: Security Goals}) and achieves seamless user mobility in and between regions to cope with SCN in 5G network. This paper proposes a region-based HO scheme for SCN. This is the first to achieve secure, privacy-preserving inter-region HO for roaming users in 5G without any additional infrastructural support (such as blockchain).
The major contributions of this paper
can be summarized as follows:

\begin{itemize}
    \item A concrete solution for SCN roaming environments in 5G, that provides a secure HO scheme supporting seamless user mobility in and between regions. To the best of our knowledge, this is the \emph{first} solution to achieve secure, privacy-preserving Inter-region HO for roaming users in 5G;
    \item An effective user membership revocation scheme for a large number of users in 5G using dynamic universal accumulator \cite{li_universal_2007};
    \item A rigorous formal security analysis of our
proposed scheme, showing that our scheme achieves mutual authentication, user unlinkability and secure key exchange.
    \item A comparative study of the proposed scheme with closely related existing schemes, showing that our scheme is secure and computationally efficient.
\end{itemize}

\subsection{Paper organisation:} The rest of the paper is organised as follows: Section \ref{Preliminaries} presents the preliminaries used in the proposed scheme. Section \ref{sec: sys and adv} introduces the system and adversary models. Section \ref{Proposed scheme} presents the proposed secure inter-region HO authentication scheme, with a detailed description of the proposed protocols, including registration, initial authentication, intra-region and inter-region HO protocols. Section\ref{Security Analysis} provides a formal security analysis of our proposed protocols. A performance evaluation and comparison of the proposed scheme with the other related schemes is presented in Section \ref{sec: Performance}. Finally, section \ref{Conclusion} concludes the paper. We give our notation used in this paper in Table \ref{tab:Notation}.
\begin{table}
    \centering
    \caption{Notation used in our proposed scheme.}
            \label{tab:Notation}
    \begin{tabular}{|l|l|}
    \hline
        \textbf{Notation} & \textbf{Meaning} \\
        \hline \hline
        EID & gNB Identity \\ 
        \hline
        ZUID & Zone user ID \\
       \hline
        HID & HgNB Identity \\ 
        \hline
        RU-ID & Region user ID \\
       \hline
        RID & Region Identity \\ 
        \hline
        $\pi_{U}$ & non-membership witness \\
       \hline
        PID,TID & User pseudo IDs \\ 
        \hline
        $T_{U}$ & User subscription validity period\\
        \hline
        $RL_{v}\,RL_{new}$ & Revocation list \\
       \hline 
        $k_{i}$ & Long-term key \\ 
        \hline
        $k_s$ & Session key \\
       \hline 
       $\AuthEnc.\Enc\{k_i,m\}$ & Authenticated encryption \\ 
        \hline
       $\AuthEnc.\Dec\{k_i,m\}$ &Authenticated decryption\\
       \hline
       $\cert_{H}$ & HgNB certificate \\ 
        \hline
       $\cert_{U}$ & UE certificate\\
        \hline
       $\pk_{sig}^{\AuC}, \sk_{sig}^{\AuC}$ & $\AuC$ public and secret signing keys\\
       \hline
        $\pk_{san}^{\gNB}, \sk_{san}^{\gNB}$ & $\gNB$ public and secret sanitising keys\\
       \hline
        $\pk_{san}^{\HgNB}, \sk_{san}^{\HgNB}$ & $\HgNB$ public and secret sanitising keys\\
       \hline
    \end{tabular}
\end{table}

\section{Preliminaries} 
\label{Preliminaries}
In this section, we introduce the underlying cryptographic primitives that we require for building the proposed protocols, in particular Sanitisable signatures and accumulators.
\begin{enumerate}[leftmargin=*]
\item \textbf{Sanitisable Signatures:}
A fundamental component of the proposed scheme are Sanitisable Signatures (SanSigs) \cite{ateniese_sanitizable_2005}, a signature scheme where signing capabilities can be delegated to another party: the so-called sanitiser. The sanitiser can modify parts of a signed message (generating another valid signature over the modified message) without the original signer's assistance. SanSig requires a pair deterministic functions $\adm$ and $\mod$, which can indicate if the message was modified correctly, or recover the original message from the modified message to its original message respectively, i.e., $m^{*}=\mod (m)$ and $\adm (m^{*},m)\to \{0,1\}$.
Sanitisable Signatures should satisfy number of  security properties: Unforgeability, Immutability, Privacy, Transparency and Accountability. A SanSig is a tuple of algorithms $\SanSig = \{\KeyGen, \Sign, \Sanit,\Verify, \proof, \Judge\}$.
However, since $\proof$ and $\Judge$ are not used in our scheme directly, we omit their description here:
\begin{itemize}
    \item $\boldsymbol{\KeyGen}$ is a pair of key generation algorithms for the signer and the sanitiser respectively: $(\pk_{sig},\sk_{sig}) \getsr \KeyGen_{sig}(1^{n})$, 
    $(\pk_{san},\sk_{san}) \getsr \KeyGen_{san}(1^{n})$.
    \item $\boldsymbol\Sign$ takes as input a message $m \in \bits{*}$, a signer private key $\sk_{sig}$, sanitiser public key $\pk_{san}$ and the modifiable message segments ($\adm$). $\Sign$ either outputs a signature $\sigma$, or $\bot$ if failed:
    $ \sigma \getsr \Sign(m,\sk_{sig}, \pk_{san}, \adm )$.
    
    \item $\boldsymbol\Sanit$ takes as input an original message $m$, a modification of the original message $\mod$, a signature $\sigma$, signer public key $\pk_{sig}$ and sanitiser private key $\sk_{san}$. $\Sanit$ outputs either a modified message $m^{*}$ and a signature $\sigma^{*}$, or $\bot$ if failed: $(m^{*},\sigma^{*}) \getsr \Sanit(m,\mod, \sigma,  \pk_{sig}, \sk_{san} )$.
    
    \item $\boldsymbol\Verify$ takes as input a message $m$, a signature $\sigma$ and the public keys of the signer $\pk_{sig}$ and sanitiser $\pk_{san}$. $\Verify$ outputs a bit $b \in \bits{}$, where $b=1$ if $\sigma$ verifies message $m$ under $\pk_{san}$ and $\pk_{sig}$, and $b=0$ otherwise:
    $b \leftarrow \mathit{\Verify}(m, \sigma, \pk_{sig}, \pk_{san} )$.
\end{itemize}

\item \textbf{Accumulator:}
\label{accu:def}

To manage the significant number of connected devices in the network, the proposed system model supports a revocation mechanism that is based on Li, Li, and Xue's dynamic universal accumulator \cite{li_universal_2007}. Their framework provides a non-membership witness for users not in the accumulator. Thus, users must prove their non-membership of the revocation list before accessing the network. Typically in mobile communication networks, the number of joining users is higher than revoked users. Therefore, using an accumulator scheme that supports non-membership witnesses is more efficient than other accumulators. The frequency of updating the accumulator (and witnesses) is less than other accumulators since it is relatively correlated to the number of revoked users, not the joining users.

\begin{itemize}
    \item \textbf {Key Generation:}
    $\KGenacc$ generates a secret key $\skacc$. 
    $[(\skacc) \getsr \KGenacc(1^{n})]$.
    
    \item \textbf {Accumulator Generation:}
    $\Genacc$ takes as input a secret key $\skacc$, and the set of values to be accumulated $X$ (Revocation list), which upon initialisation $X \gets \phi$. It returns an accumulator $acc$.
    $[(acc) \getsr \Genacc(\skacc,X)]$.

    \item \textbf {Accumulator Update:} $\Updateacc$ takes as input a secret key $\skacc$, the original accumulator value $acc$ and a new value to be accumulated $x^{*}$. It returns the updated accumulator $acc^{*}$.
    $[(acc^{*}) \getsr \Updateacc(\skacc,acc,x^{*} )]$.

    \item \textbf {Non-membership Witness Generation:}
     $\Nwitcreate$ takes as input a secret key $\skacc$, the original accumulator $acc$, the revocation list $X$ and $x^{*}$, where $x^{*} \notin X$. It returns a non-membership witness $c_{x}$ for $x^{*}$.
    $[(c_{x}) \getsr \Nwitcreate(\skacc,acc,X, x^{*} )]$.     
    
    \item \textbf {Non-membership Witness Verification:}
    $\Verifyacc$ takes as input the original accumulator value $acc$, a non-membership witness $c_{x}$ and $x$. It outputs a bit $b \in \bits{}$, where $b=1$ if witness $c_{x}$ holds (hence the value $x \notin X$), and $b=0$ otherwise.
    $[b \leftarrow \mathit \Verifyacc(acc,c_{x},x )]$.

    \item \textbf {Non-membership Witness Update}
    $\Nwitupdate$ takes as input the original accumulator $acc$, the updated accumulator $acc^{*}$, a (new) accumulated value $x^{*}$, an non-accumulated value $x$ and the original non-membership witness $c_{x}$. It outputs an updated non-membership witness $c^{*}_{x}$. This is required if a new element $x^{*}$ has been added to the accumulator $acc^{*}$.
    
    $[(c^{*}_{x}) \getsr \Nwitupdate(acc,acc^{*},x^{*},x, c_{x})]$ 
  
    \end{itemize}

\end{enumerate}

\section{System and Adversary model}
\label{sec: sys and adv}
In this section, we first describe the system model and adversary model of our proposed schemes. Subsequently, we define the security goals of the proposed scheme.

\subsection{System model}

\begin{figure*}
    \centering
    \includegraphics [scale=0.545]{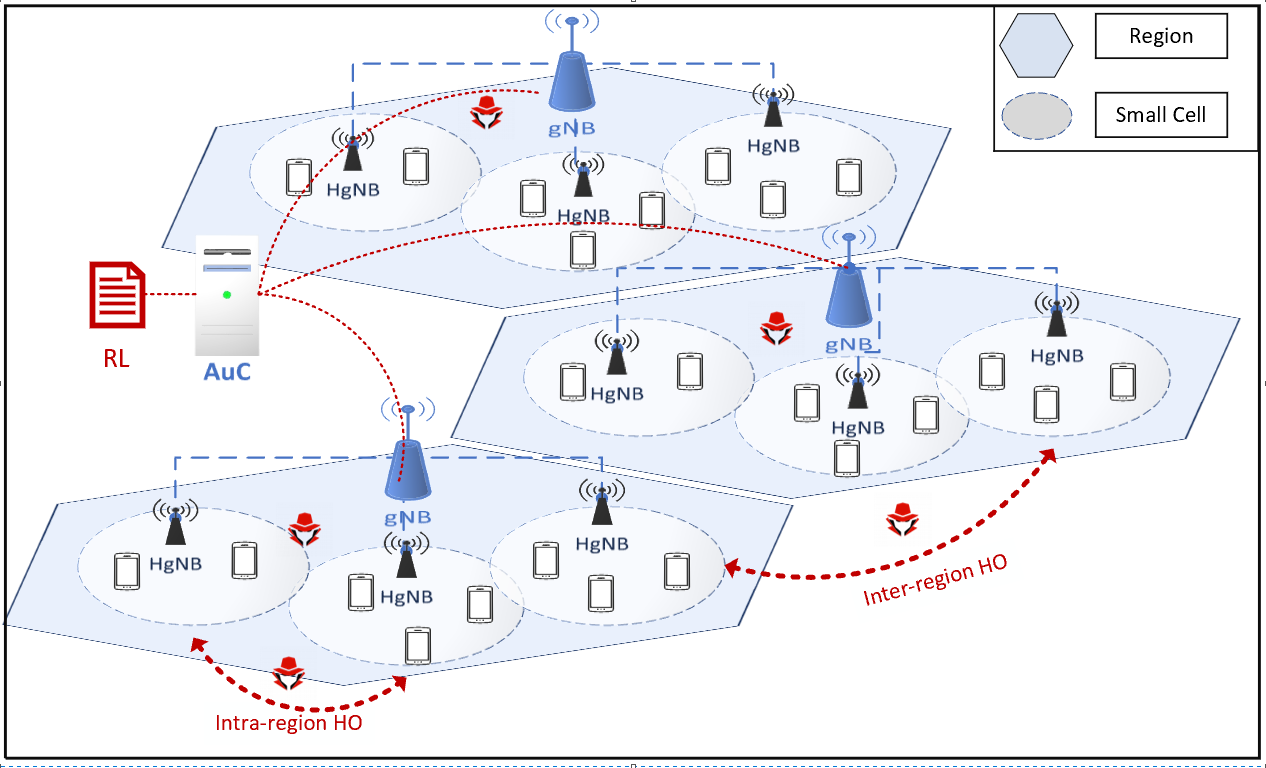}
    \caption{System model [Hexagon shape $\rightarrow$ a region managed by one gNB, Oval shape $\rightarrow$ small cell emulating SCN.]}
    \label{fig:Sys model}

\end{figure*}

{Our system model captures the architecture of SCNs in 5G~\cite{3gpp_etsi_2018}, which consists of four major components; the Authentication Center (AuC), the 5G radio base station that connects users to the 5G core network gNB, Home gNB (HgNB) and User Equipment (UE). In SCN, HgNB management System (HeMS) is responsible for configuring the gNB/HgNB according to the operators policy. In our system model, however, we combine HeMS with the AuC. Thus the AuC is responsible for configuring all parties in the system including the HgNB, gNB and UE, where AuC generates certificates, secret and public keys for them. Meanwhile, gNB and HgNB are responsible for connecting users to the core network. Each gNB in our system model manages a group of several HgNBs, creating a \emph{Region} controlled by an gNB, as illustrated in Figure \ref{fig:Sys model}. As a result, the gNB is responsible for handling the Inter-region HOs, while HgNB is responsible for handling the Intra-region HOs and key exchange. It is assumed that the communication channels between network entities i.e., AuC, gNB and HgNB are secure, i.e. an authenticated and confidential channel. }

Note that for the standard 5G handover scheme \cite{peltonen_comprehensive_2021}, the cooperation of AuC (i.e., all 5G core entities) is inevitable. That means the AuC needs to be actively involved during the execution of each handover phase. This will increase the transmission overhead on AuC. Accordingly, it will also increase the required time to perform a handover and affect the security and user privacy. This issue will be exacerbated significantly with the inclusion of SCN in the 5G networks. Hence, in the proposed scheme, each \emph{Region} consists of a gNB and its belonging HgNBs. The legitimacy of users roaming inside a region can be verified by a designated HgNB using the public key of the gNB of the same Region. Meanwhile, users roaming between regions are verified, and their certificates are modified (to preserve user anonymity) by the gNB of the targeted Region (as a sanitiser). Thus, in our proposed system, we do not require the 5G core/AuC to be actively involved or participated during the execution of the handover phases (intra-region and inter-region).
\subsection{Adversary Model}
In our system model, a user communicates with other network entities, i.e., HgNBs and gNBs, through the public/insecure channels. Hence, in our adversary model, we allow an adversary $\adv$ to control the public channels fully; therefore, $\adv$ can intercept, delete, insert and modify any message. Finally, $\adv$ can leak long-term and per-session secrets, capturing device-compromising attackers. 
User privacy is considered an essential property in 5G-based mobile communication, so $\adv$ may also try to break User anonymity by linking distinct ``challenge'' protocol executions of the same user, capturing linkability attacks. To strengthen our adversary model, we allow $\adv$ to schedule session initialisation, thus all sessions are initialised with owners chosen by the adversary, except the ``challenge'' sessions, which are instead initialised with a pair of potential owners, i.e. $\adv$ must distinguish which party owns the ``challenge'' session. In addition, we wish to ensure that an adversary cannot break authentication, nor learn session keys established during the proposed handover schemes. In Section \ref{Security Analysis}, we present security experiments capturing these notions.

\subsection{Security Goals}
\label{subsec: Security Goals}
Our proposed privacy-preserving handover protocols should achieve the following security goals: 

\emph{\textbf{Mutual authentication:}}
It is essential to guarantee the authenticity of the communicating parties for both the network components and the UEs. Otherwise, this may cause various security threats such as MITM and impersonation attacks. We provide details of this security experiment in Section~\ref{sec:MA-game}.

 \emph{\textbf{Strong anonymity:}} Privacy is one of the major security requirements in 5G, where users' identities should never be transmitted in plaintext. However, encryption alone is not sufficient for preserving user privacy, as user identities can be revealed by honest-but-curious network components. To address this, our schemes replace users' long-term identities with (encrypted) temporary identifiers. Providing user anonymity is not sufficient for fulfilling the privacy requirements in 5G. Although users' identities are anonymised, attackers could link distinct sessions to a single user, causing traceability attacks. Our scheme achieves security against traceability attacks even in the SCN, where $\adv$ can monitor network traffic between the small cells. We provide details of the security experiment in Section \ref{sec:unlink-game}.

\emph{\textbf{Perfect forward secrecy:}}
Ensuring the security of previously-computed session keys after long-term secrets are compromised is essential for 5G. Ephemeral Diffie-Hellman key exchange (authenticated via sanitisable signatures) allows our proposed scheme to achieve this notion. Use of  SCNs increases the frequency of HO protocol executions, which shortens the window of session key compromise. We provide details of this security experiment in Section \ref{sec:Key Ind-game}.

\emph{\textbf{Effective revocation management:}}
Since the number of joining users to the 5G network is continuously increasing, providing a subscription management mechanism is essential. Therefore, we utilise the universal accumulators for our proposed protocol's revocation list (RL).

\section{Secure Region-based Handover Scheme}
\label{Proposed scheme}

Here we introduce our proposed handover scheme consisting of \emph{four} phases: \emph{Registration}, \emph{Initial authentication}, \emph{Intra-region HO}, and \emph{Inter-region HO}.

The \textit{Registration} phase registers gNBs, HgNBs and new users in the network, generating the initialisation parameters for the network parties (AuC, gNB and HgNB) (where AuC creates $\SanSig$ key pairs for gNBs and certificates for HgNB). Additionally, AuC generates and shares a long-term secret key with the User and pseudo-identities (PIDs \& TIDs) for user anonymity. Then $\AuC$ initialises the user membership revocation list (RL), which is initially empty. 
The \textit{Initial Authentication} phase issues certificates for new users in the network (via $\SanSig$) using their PIDs. 
The \textit{Intra-Region HO} phase allows users to roam inside a region (between two HgNBs owned by a single gNB) to mutually authenticate the target HgNB using their certificates', and establish a shared secret key. 
The \textit{Inter-Region HO} phase allows  users to roam between regions (between two HgNBs controlled by different gNBs) to mutually authenticate the target gNB and establish a shared secret key.
\subsection{Registration Phase}

\iffullversion
In the registration phase of the proposed scheme, we assume that the communication channels between the parties (i.e. AuC, gNB, HgNB, UE) are secure. During the execution of the registration phase, AuC generates the required credentials for all participants, including the HgNB certificate, HgNB sanitising keys, gNB sanitising keys, UE pseudo-identities ($TID,PID$), UE long term key ($k_i$) and the revocation list (RL). Our registration phase is divided into three parts: $\UE$ registration, $\gNB/\HgNB$ registration, and accumulator initialisation, which are described as follows.

\fi

    \begin{enumerate}[leftmargin=*]
\item \textbf{$\boldsymbol\UE$ Registration}: In order to register into the network, each $\UE_i$ needs to share his/her essential information with the $\AuC$ via a secure channel. Upon receiving the registration request, the $\AuC$ then generates a long-term secret key $k_{i}$, a pseudo-identity ($\PID_{i}$) and a temporary ID ($\TID_{i}$) for each user, where $\PID_{i},\TID_{i} \getsr \bits{n}$. 

\item \textbf{$\boldsymbol{\gNB/\HgNB}$ Registration}:
Each $\gNB$ and $\HgNB$ needs to register to the network and share the essential registration information with the $\AuC$. Upon receiving the registration request, the $\AuC$ then generates a signing key pair for itself and HgNBs i.e., $(\pk_{sig}^{\AuC}, \sk_{sig}^{\AuC}) \getsr \SanSig.\KeyGen_{sig}(1^{n}) $,
$(\pk_{san}^{\HgNB}, \sk_{san}^{\HgNB}) \getsr \SanSig.\KeyGen_{san}(1^{n})$.
Next, AuC signs the $\HgNB_i$ certificate ($\cert_{H}= \cert_{fix}^{H}||\cert_{mod}^{H}$) for each HgNB in the network: $\sigma_{H} \getsr \SanSig.\Sign(\cert_{H}, \sk_{sig}^{\AuC}, \pk_{san}^{\HgNB}, \\ \adm(\cert_{mod}^{H}) )$. Hereafter, AuC generates signing and sanitising keys for itself and gNBs: $
(\pk_{sig}^{\AuC}, \sk_{sig}^{\AuC}) \getsr \SanSig.\KeyGen_{sig}(1^{n}) $,$(\pk_{san}^{\gNB},\sk_{san}^{\gNB})\getsr \SanSig.\KeyGen_{san}(1^{n})$. These pairs of keys will be used to sign users' certificates ($\cert_{U}$) in the initial authentication phase, and sanitise users' certificates in the inter-region phase. To expedite the registration process, AuC can execute this step offline.

\item\textbf{Accumulator Initialisation}: To initialise the accumulator/revocation list, first the AuC generates a secret accumulator key $(\skacc) \getsr \KGenacc(1^{n})$ and also creates the revocation list $\RL \getsr \Genacc(\skacc,X)$, where $X$ is initially empty.
    \end{enumerate}

\subsection{Initial authentication}

\iffullversion
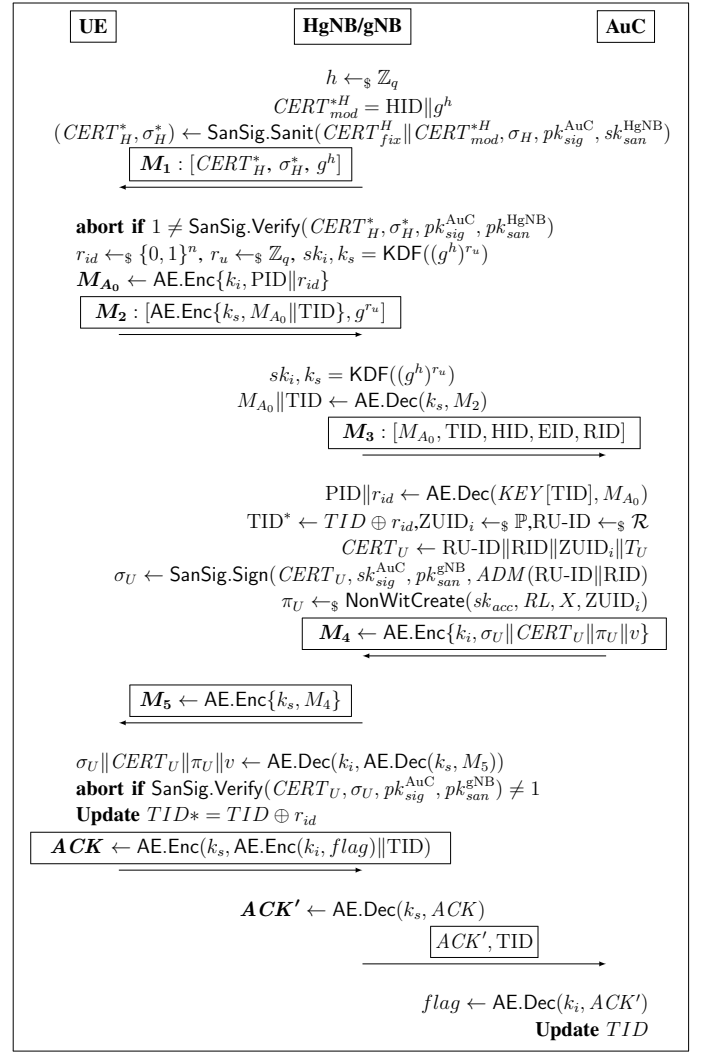
\begin{figure}[t!]
	\centering
	\begin{adjustbox}{max width=0.5\textwidth, max height=1\textheight}
		\fbox{
	\begin{tikzpicture}[yscale=-0.55,>=latex]
    \tikzstyle{every node}=[font=\large]
	\edef\InitX{0}
	\edef\ArrowLeft{1}
	\edef\ArrowCenter{6}
	\edef\ArrowRight{11}
	\edef\RespX{12}
	\edef\Y{0}
	
	\node [rectangle,draw,inner sep=5pt,right] at (\InitX,\Y) {\textbf{UE}};
	\node [rectangle,draw,inner sep=5pt,left] at ((7,\Y) {\textbf{HgNB/gNB}};
	\node [rectangle,draw,inner sep=5pt,left] at (\RespX,\Y) {\textbf{AuC}};
	\NextLine[2]
	\AdversaryAction{$h \getsr \mathbb{Z}_q$}
	\NextLine
	\AdversaryAction{$\cert^{*H}_{mod}=\HID\|g^{h}$}
	\NextLine
	\AdversaryAction{$(\cert_{H}^{*},\sigma^{*}_{H}) \gets \SanSig.\Sanit (\cert^{H}_{fix}\|\cert^{*H}_{mod},\sigma_{H}, \pk_{sig}^{\AuC}, \sk_{san}^{\HgNB} )$}
	\NextLine[1.5]
	\AdversaryToClient{\framebox[1.1\width]{$\boldsymbol{M_1}:[\cert_{H}^{*}$, $\sigma^{*}_{H}$, $g^h]$}}{}
	\NextLine
	\ClientAction{\textbf{abort if} $1 \neq \SanSig.\Verify(\cert_{H}^{*}, \sigma^{*}_{H}, \pk_{sig}^{\AuC}, \pk_{san}^{\HgNB} )$}
	\NextLine
	\ClientAction{$r_{id} \getsr \bits{n}$, $r_{u} \getsr \mathbb{Z}_q$, $sk_{i},k_s=\KDF((g^h)^{r_{u}})$}
	\NextLine
	\ClientAction{$\boldsymbol{M_{A_{0}}} \gets \AuthEnc.\Enc\{k_{i},\PID\|r_{id}\}$}
	\NextLine[1.5]
	\ClientToAdversary{\framebox[1.1\width]{$\boldsymbol{M_2}:[\AuthEnc.\Enc\{k_s,M_{A_{0}}\|\TID\},g^{r_u}]$}}{}
	\NextLine
	\AdversaryAction{$sk_{i},k_s=\KDF((g^h)^{r_{u}})$}
	\NextLine
	\AdversaryAction{$M_{A_{0}}\|\TID \gets \AuthEnc.\Dec(k_s,M_{2})$}
	\NextLine[1.5]
	\AdversaryToServer{\framebox[1.1\width]{$\boldsymbol{M_3}:[M_{A_{0}},\TID,\HID,\EID,\RID]$}}{}
	\NextLine
	\ServerAction{$\PID\|r_{id} \gets \AuthEnc.\Dec(\mathit{KEY}[\TID],M_{A_{0}})$}
	\NextLine
	\ServerAction{$\TID^*\gets TID \oplus r_{id}$,$\ZUID_i\getsr \mathbb{P}$,$\RUID \getsr \mathcal{R}$}
	\NextLine
	\ServerAction{$\mathit{CERT}_U \gets \RUID\|\RID\|\ZUID_i\|T_U$}
	\NextLine
	\ServerAction{$\sigma_U \gets \SanSig.\Sign(\mathit{CERT}_U,\sk_{sig}^{\AuC},\pk_{san}^{\gNB},\mathit{ADM}(\RUID\|\RID)$}
	\NextLine
	\ServerAction{$\pi_{U} \getsr \mathsf{NonWitCreate}(\sk_{acc},\mathit{RL},X,\ZUID_i)$}
	\NextLine[1.5]
	\ServerToAdversary{\framebox[1.1\width]{$\boldsymbol{M_{4}}\gets\AuthEnc.\Enc\{k_{i},\sigma_{U}\|\cert_{U}\|\pi_{U} \|v\}$}}{}
	\NextLine[1.5]
	\AdversaryToClient{\framebox[1.1\width]{$\boldsymbol{M_{5}}\gets\AuthEnc.\Enc\{k_s,M_{4}\}$}}{}
	\NextLine
	\ClientAction{$\sigma_{U}\|\cert_{U}\|\pi_{U} \|v \gets \AuthEnc.\Dec(k_i,\AuthEnc.\Dec(k_s,M_{5}))$}
	\NextLine
	\ClientAction{\textbf{abort if} $\SanSig.\Verify(\cert_{U},\sigma_{U},\pk_{sig}^{\AuC},\pk_{san}^{\gNB})\neq1$}
	\NextLine
	\ClientAction{\textbf{Update}
	$TID*=TID \oplus r_{id}$}
	\NextLine[1.5]
	\ClientToAdversary{\framebox[1.1\width]{$\boldsymbol{\mathit{ACK}} \gets \AuthEnc.\Enc(k_s,\AuthEnc.\Enc(k_i,flag)\|\TID)$}}{}
	\NextLine
	\AdversaryAction{$\boldsymbol{\mathit{ACK}'} \gets \AuthEnc.\Dec(k_s,\mathit{ACK})$}
	\NextLine[1.5]
	\AdversaryToServer{\framebox[1.1\width]{$\mathit{ACK}',\TID$}}{}
	\NextLine
	\ServerAction{$flag \gets \AuthEnc.\Dec(k_i,\mathit{ACK'})$}
	\NextLine
    \ServerAction{\textbf{Update} $TID$}
	\end{tikzpicture}
}
	\end{adjustbox}
	\caption{The Initial Authentication Phase of our proposed 5G Secure Handover Scheme. The descriptions of each algorithm can be found in Section \ref{Preliminaries}.}
	\label{fig:init-auth-alt}
\end{figure}
\else
\begin{figure}[htb]
	\centering
	\begin{adjustbox}{max width=1\textwidth, max height=1\textheight}
		\fbox{
	\begin{tikzpicture}[yscale=-0.55,>=latex]
    \tikzstyle{every node}=[font=\large]
	\edef\InitX{0}
	\edef\ArrowLeft{1}
	\edef\ArrowCenter{6}
	\edef\ArrowRight{11}
	\edef\RespX{12}
	\edef\Y{0}
	
	\node [rectangle,draw,inner sep=5pt,right] at (\InitX,\Y) {\textbf{UE}};
	\node [rectangle,draw,inner sep=5pt,left] at ((7,\Y) {\textbf{HgNB/gNB}};
	\node [rectangle,draw,inner sep=5pt,left] at (\RespX,\Y) {\textbf{AuC}};
	\NextLine[2]
	\AdversaryAction{$h \getsr \mathbb{Z}_q$}
	\NextLine
	\AdversaryAction{$\cert^{*H}_{mod}=\HID\|g^{h}$}
	\NextLine
	\AdversaryAction{$(\cert_{H}^{*},\sigma^{*}_{H}) \gets \SanSig.\Sanit (\cert^{H}_{fix}\|\cert^{*H}_{mod},\sigma_{H}, \pk_{sig}^{\AuC}, \sk_{san}^{\HgNB} )$}
	\NextLine[1.5]
	\AdversaryToClient{\framebox[1.1\width]{$\boldsymbol{M_1}:[\cert_{H}^{*}$, $\sigma^{*}_{H}$, $g^h]$}}{}
	\NextLine
	\ClientAction{\textbf{abort if} $1 \neq \SanSig.\Verify(\cert_{H}^{*}, \sigma^{*}_{H}, \pk_{sig}^{\AuC}, \pk_{san}^{\HgNB} )$}
	\NextLine
	\ClientAction{$r_{id} \getsr \bits{n}$, $r_{u} \getsr \mathbb{Z}_q$, $sk_{i},k_s=\KDF((g^h)^{r_{u}})$}
	\NextLine
	\ClientAction{$\boldsymbol{M_{A_{0}}} \gets \AuthEnc.\Enc\{k_{i},\PID\|r_{id}\}$}
	\NextLine[1.5]
	\ClientToAdversary{\framebox[1.1\width]{$\boldsymbol{M_2}:[\AuthEnc.\Enc\{k_s,M_{A_{0}}\|\TID\},g^{r_u}]$}}{}
	\NextLine
	\AdversaryAction{$sk_{i},k_s=\KDF((g^h)^{r_{u}})$}
	\NextLine
	\AdversaryAction{$M_{A_{0}}\|\TID \gets \AuthEnc.\Dec(k_s,M_{2})$}
	\NextLine[1.5]
	\AdversaryToServer{\framebox[1.1\width]{$\boldsymbol{M_3}:[M_{A_{0}},\TID,\HID,\EID,\RID]$}}{}
	\NextLine
	\ServerAction{$\PID\|r_{id} \gets \AuthEnc.\Dec(\mathit{KEY}[\TID],M_{A_{0}})$}
	\NextLine
	\ServerAction{$\TID^*\gets TID \oplus r_{id}$,$\ZUID_i\getsr \mathbb{P}$,$\RUID \getsr \mathcal{R}$}
	\NextLine
	\ServerAction{$\mathit{CERT}_U \gets \RUID\|\RID\|\ZUID_i\|T_U$}
	\NextLine
	\ServerAction{$\sigma_U \gets \SanSig.\Sign(\mathit{CERT}_U,\sk_{sig}^{\AuC},\pk_{san}^{\gNB},\mathit{ADM}(\RUID\|\RID)$}
	\NextLine
	\ServerAction{$\pi_{U} \getsr \mathsf{NonWitCreate}(\sk_{acc},\mathit{RL},X,\ZUID_i)$}
	\NextLine[1.5]
	\ServerToAdversary{\framebox[1.1\width]{$\boldsymbol{M_{4}}\gets\AuthEnc.\Enc\{k_{i},\sigma_{U}\|\cert_{U}\|\pi_{U} \|v\}$}}{}
	\NextLine[1.5]
	\AdversaryToClient{\framebox[1.1\width]{$\boldsymbol{M_{5}}\gets\AuthEnc.\Enc\{k_s,M_{4}\}$}}{}
	\NextLine
	\ClientAction{$\sigma_{U}\|\cert_{U}\|\pi_{U} \|v \gets \AuthEnc.\Dec(k_i,\AuthEnc.\Dec(k_s,M_{5}))$}
	\NextLine
	\ClientAction{\textbf{abort if} $\SanSig.\Verify(\cert_{U},\sigma_{U},\pk_{sig}^{\AuC},\pk_{san}^{\gNB})\neq1$}
	\NextLine
	\ClientAction{\textbf{Update}
	$TID*=TID \oplus r_{id}$}
	\NextLine[1.5]
	\ClientToAdversary{\framebox[1.1\width]{$\boldsymbol{\mathit{ACK}} \gets \AuthEnc.\Enc(k_s,\AuthEnc.\Enc(k_i,flag)\|\TID)$}}{}
	\NextLine
	\AdversaryAction{$\boldsymbol{\mathit{ACK}'} \gets \AuthEnc.\Dec(k_s,\mathit{ACK})$}
	\NextLine[1.5]
	\AdversaryToServer{\framebox[1.1\width]{$\mathit{ACK}',\TID$}}{}
	\NextLine
	\ServerAction{$flag \gets \AuthEnc.\Dec(k_i,\mathit{ACK'})$}
	\NextLine
    \ServerAction{\textbf{Update} $TID$}
	\end{tikzpicture}
}
	\end{adjustbox}
	\caption{The Initial Authentication Protocol of our proposed 5G Secure Handover Scheme. The descriptions of each algorithm can be found in Section \ref{Preliminaries}.}
	\label{fig:init-auth-alt}
\end{figure}
\fi

\iffullversion
Each registered user who wants to join the network needs to participate in the initial authentication phase of our proposed scheme. During the execution of this protocol, the AuC generates credentials/certificates for new users, which will be used in the subsequent phases (Handover protocols). In this protocol, $\gNB$s are passive entities, where they are only responsible for forwarding messages between $\HgNB$s and $\UE$s. Therefore, we combine the $\HgNB$ and $\gNB$ into one entity in this protocol. This protocol consists of the following steps and is illustrated in Figure \ref{fig:init-auth-alt}.
\fi

\begin{itemize}[leftmargin=*]

    \item[]\textbf{Step $\mathbf{A_1}$:} $\HgNB \rightarrow \UE$: $\mathbf{M_{1}}$: [$\cert_{H}^{*},\sigma^{*}_{H},g^{h}$].\\  
    When a new $\UE$ enters the coverage area of HgNB, HgNB samples $h$ and computes $g^{h}$. Next, HgNB updates their certificate (the modifiable part) , i.e. $\cert^{*H}_{mod}=\HID\|g^{h}$ (preventing replay attacks). Then $\HgNB$ sanitises the updated certificate $\cert_{H}=\cert^{H}_{fix}\|\cert^{*H}_{mod}$, using the sanitising algorithm $\SanSig.\Sanit$, and composes a message $\mathbf{M_{1}}$, sending $\mathbf{M_{1}}$ to UE.
   
\item[]\textbf{Step $\mathbf{A_2}$:}$\UE \rightarrow \HgNB/\gNB$: \\$\mathbf{M_{2}}$:$[ \AuthEnc.\Enc\{k_s,M_{A_{0}}||\TID\},g^{r_{u}}]$.

Upon receiving $\mathbf{M_{1}}$, UE verifies the $\HgNB$ certificate $\cert_{H}$ using the $\SanSig$ verification algorithm $\SanSig.\Verify$, containing $g^{h}$ (preventing MITM attacks). If successful, UE samples $(r_{id}, r_{u})$, and computes the session keys $(sk_{i},k_s)$. Next, UE encrypts  $(\PID||r_{id})$ using the long-term key $k_{i}$ shared with AuC, to generate the message $\mathbf{M_{A_0}}$. Afterwards, UE encrypts $(M_{A_{0}}||\TID)$ with $k_s$ (preventing linkability), and sends the message $\mathbf{M_{2}}$ to HgNB.

\item[]\textbf{Step $\mathbf{A_3}$:} $\HgNB/\gNB \rightarrow \AuC$:   \\$\mathbf{M_{3}}:[M_{A_{0}},\TID, \HID, \EID, \RID].$\\
Upon receiving the response message $\mathbf{M_{2}}$, HgNB computes $(sk_{i},k_s)$ to decrypt $\mathbf{M_{2}}$. Next HgNB forwards the decrypted message along with the user pseudo-identities and region identities to AuC.

\item[]\textbf{Step $\mathbf{A_4}$:} $\AuC \rightarrow \HgNB/\gNB$: \\$\mathbf{M_{4}}: [\AuthEnc. \Enc\{k_i,\sigma_{U}\|\cert_{U}\|\pi_{U}\|v)\}].$\\
AuC retrieves the long term key $k_i$  of UE using $\TID$, and decrypts $\mathbf{M_{A_0}}$, to recover $(\PID, r_{id}) $.
Next, AuC computes a new temporary user identifier $\TID^*$, and generates a user ID $(\ZUID_i)$, which will be the user's identifier in the revocation list $\RL$. AuC creates and signs $\cert_{U}$ by generating the ``fixed'' part of the UE certificate $\cert_{fix}^{U} = \ZUID_i \| T_{U}$ (where $T_{U}$ is a user subscription validity period), and the ``modifiable'' region-specific part of the UE certificate $\cert_{mod}^{U} = \RUID\|\RID$ (where $\RUID$ a region-user ID and $\RID$ is the region ID). Then AuC signs both parts of the UE certificate generating $\cert_U \gets \SanSig.\Sign$.
Next, AuC generates a non-membership witness $(\pi_{U})\gets \Nwitcreate$, and specifies the version $v$ of $\RL$, corresponding to the version of $\RL$ from which $\pi_{U}$ was generated. AuC then stores $\ZUID_i$, $\TID_i$ and $\TID^*_i$ (to prevent de-synchronisation), and encrypts $\pi_{U}$, UE certificate and its signature using $k_i$, to generate the message $\mathbf{M_{4}}$ sending $\mathbf{M_{4}}$ to the HgNB/gNB.

\item[]\textbf{Step $\mathbf{A_5}$:}  $\HgNB \rightarrow \UE$: $\mathbf{M_{5}} :[ \AuthEnc.\Enc\{k_s,M_{4}\}]$

HgNB/gNB encrypts $\mathbf{M_{4}}$ using the session key $k_s$ (preventing linkability) to generate the message $\mathbf{M_{5}}$, sending $\mathbf{M_{5}}$ to the UE.

\item[]\textbf{Step $\mathbf{A_6}$:} $\UE \rightarrow \AuC$: $\mathbf{ACK:} [\AuthEnc.\Enc(k_s,$ $(\AuthEnc.\Enc(k_i,flag),\TID))]$

Upon receiving $\mathbf{M_{5}}$, UE recovers $(\sigma_{U},CERT_{U},\pi_{U},v) $, and verifies their certificate, using $\SanSig.\Verify(\cert_{U},\sigma_{U})$. If verification fails, $\UE$ terminates the execution of the protocol. Otherwise, the user then updates $\TID^*$, and sends an acknowledgement $\mathbf{ACK}$, an encryption of an acknowledgement flag $flag$ and the user's $\TID$, which is encrypted using the user long term key $k_i$ and $\TID$, and then encrypted again using the ephemeral key $k_s$ to HgNB.

\item[]\textbf{Step $\mathbf{A_7}$:}  $\HgNB \rightarrow \AuC$: $ACK' :[ ACK',\TID]$

HgNB/gNB decrypts $\mathbf{ACK}$ using the session key $k_s$ (preventing linkability) to generate the message $ACK'$, sending $ACK',\TID$ to the AuC.

\item[]\textbf{Step $\mathbf{A_8}$:} Upon receiving $ACK$, AuC recovers $k_i$ (using the old TID), and uses it to decrypt $ACK$, then Auc updates the $TID^*$. If $ACK$ was not received within the pre-specified time window, $\AuC$ deletes $\TID^*$. $AuC$ will continue to maintain both $\TID$ and $\TID^*$. Details of this protocol is depicted in Figure 2.

\begin{remark}
   To prevent de-synchronisation (DoS attacks)\cite{gope_resilience_2017}
   AuC maintains both ($\TID$, $\TID^*$) values until receiving the $ACK$ message. However, to prevent DoS attacks without compromising the privacy of the UE, we can use the previous solution proposed in  \cite{gope_laap_2019} that overcomes de-synchronisation attacks by utilising a family of temporary PIDs instead of a single TID.
    \end{remark}

\end{itemize}

\subsection{Intra-region Handover}
The intra-region handover protocol will be executed when a user remains under the same region where he/she was in but roams to a different small-cell under authority of a different HgNB i.e., between HgNBs belonging to the same region. The intra-region HO protocol is described below, and illustrated in Figure \ref{fig:Intra-region-alt}.

\iffullversion
\begin{figure}[t!]
	\centering
	\begin{adjustbox}{max width=0.5\textwidth, max height=1\textheight}
		\fbox{
	\begin{tikzpicture}[yscale=-0.55,>=latex]
    \tikzstyle{every node}=[font=\large]
	\edef\InitX{0}
	\edef\ArrowLeft{1}
	\edef\ArrowCenter{6}
	\edef\ArrowRight{11}
	\edef\RespX{12}
	\edef\Y{0}
	
    \node [rectangle,draw,inner sep=6pt,right] at (\InitX,\Y) {\textbf{UE} };
	\node [rectangle,draw,inner sep=6pt,left] at (\RespX,\Y) {\textbf{HgNB}};
	\NextLine[2]
	\ServerAction{$h \getsr \mathbb{Z}_q$}
	\NextLine
	\ServerAction{$\cert^{*H}_{mod}=\HID\|g^{h}$}
	\NextLine
	\ServerAction{$(\cert_{H}^{*},\sigma^{*}_{H}) \gets \SanSig.\Sanit (  \cert^{H}_{fix}\|\cert^{*H}_{mod},\sigma_{H}, \pk_{sig}^{\AuC}, \sk_{san}^{\HgNB} )$}
	\NextLine[1.5]
	\ServerToClient{\framebox[1.1\width]{$\boldsymbol{M_1}:[\cert_{H}^{*}$, $\sigma^{*}_{H}$, $g^h]$}}{}
	\NextLine
	\ClientAction{\textbf{abort if} $1 \neq \SanSig.\Verify(\cert_{H}^{*}, \sigma^{*}_{H}, \pk_{sig}^{\AuC}, \pk_{san}^{\HgNB} )$}
	\NextLine
	\ClientAction{$r_{u} \getsr \mathbb{Z}_q$, $sk_{i},k_s=\KDF((g^h)^{r_{u}})$}
	\NextLine[1.5]
	\ClientToServer{\framebox[1.1\width]{$\boldsymbol{M_2}:[\AuthEnc.\Enc\{k_s,\cert_{U}\|\sigma_U\| \pi_U\|v\},g^{r_u}]$}}{}
	\NextLine
	\ServerAction{$sk_{i},k_s=\KDF((g^h)^{r_{u}})$}
	\NextLine
	\ServerAction{\textbf{abort if} $1 \neq \SanSig.\Verify(\cert_{U}, \sigma_{U}, \pk_{sig}^{\AuC}, \pk_{san}^{\gNB} )$}
	\NextLine
	\ServerAction{\textbf{abort if} $1 \neq \mathit \Verifyacc(RL_v,\pi_U,ZUID )$}
	\NextLine
	\ServerAction{\textbf{Update} $ [(\pi^{*}_{U}) \getsr \Nwitupdate(RL,RL^{*},x^{*},ZUID, \pi_U)]$}
	\NextLine[1.5]
   	\ServerToClient{\framebox[1.1\width]{$\boldsymbol{M_{3}}\gets\AuthEnc.\Enc\{k_s,\pi^{*}_{U}\|v*\}$}}{}
	\NextLine
	\ClientAction{$\pi^{*}_{U}\|v* \gets \AuthEnc.\Dec(k_s,M_{3})$}
	\NextLine
	\ClientAction{\textbf{Store} $(\pi^{*}_{U}\|v*)$}
	
	\end{tikzpicture}
}
	\end{adjustbox}	\caption{The Intra-region Handover Protocol of our proposed 5G Secure Handover Scheme. Descriptions of each algorithm can be found in Section \ref{Preliminaries}.}
	\label{fig:Intra-region-alt}
\end{figure}
\else
\begin{figure}[htb]
	\centering
	\begin{adjustbox}{max width=1\textwidth, max height=1\textheight}
		\fbox{
	\begin{tikzpicture}[yscale=-0.55,>=latex]
    \tikzstyle{every node}=[font=\large]
	\edef\InitX{0}
	\edef\ArrowLeft{1}
	\edef\ArrowCenter{6}
	\edef\ArrowRight{11}
	\edef\RespX{12}
	\edef\Y{0}
	
    \node [rectangle,draw,inner sep=6pt,right] at (\InitX,\Y) {\textbf{UE} };
	\node [rectangle,draw,inner sep=6pt,left] at (\RespX,\Y) {\textbf{HgNB}};
	\NextLine[2]
	\ServerAction{$h \getsr \mathbb{Z}_q$}
	\NextLine
	\ServerAction{$\cert^{*H}_{mod}=\HID\|g^{h}$}
	\NextLine
	\ServerAction{$(\cert_{H}^{*},\sigma^{*}_{H}) \gets \SanSig.\Sanit (  \cert^{H}_{fix}\|\cert^{*H}_{mod},\sigma_{H}, \pk_{sig}^{\AuC}, \sk_{san}^{\HgNB} )$}
	\NextLine[1.5]
	\ServerToClient{\framebox[1.1\width]{$\boldsymbol{M_1}:[\cert_{H}^{*}$, $\sigma^{*}_{H}$, $g^h]$}}{}
	\NextLine
	\ClientAction{\textbf{abort if} $1 \neq \SanSig.\Verify(\cert_{H}^{*}, \sigma^{*}_{H}, \pk_{sig}^{\AuC}, \pk_{san}^{\HgNB} )$}
	\NextLine
	\ClientAction{$r_{u} \getsr \mathbb{Z}_q$, $sk_{i},k_s=\KDF((g^h)^{r_{u}})$}
	\NextLine[1.5]
	\ClientToServer{\framebox[1.1\width]{$\boldsymbol{M_2}:[\AuthEnc.\Enc\{k_s,\cert_{U}\|\sigma_U\| \pi_U\|v\},g^{r_u}]$}}{}
	\NextLine
	\ServerAction{$sk_{i},k_s=\KDF((g^h)^{r_{u}})$}
	\NextLine
	\ServerAction{\textbf{abort if} $1 \neq \SanSig.\Verify(\cert_{U}, \sigma_{U}, \pk_{sig}^{\AuC}, \pk_{san}^{\gNB} )$}
	\NextLine
	\ServerAction{\textbf{abort if} $1 \neq \mathit \Verifyacc(RL_v,\pi_U,ZUID )$}
	\NextLine
	\ServerAction{\textbf{Update} $ [(\pi^{*}_{U}) \getsr \Nwitupdate(RL,RL^{*},x^{*},ZUID, \pi_U)]$}
	\NextLine[1.5]
   	\ServerToClient{\framebox[1.1\width]{$\boldsymbol{M_{3}}\gets\AuthEnc.\Enc\{k_s,\pi^{*}_{U}\|v*\}$}}{}
	\NextLine
	\ClientAction{$\pi^{*}_{U}\|v* \gets \AuthEnc.\Dec(k_s,M_{3})$}
	\NextLine
	\ClientAction{\textbf{Store} $(\pi^{*}_{U}\|v*)$}
	
	\end{tikzpicture}
}
	\end{adjustbox}
	\caption{The Intra-region Handover Protocol  of our proposed 5G Secure Handover Scheme. Descriptions of each algorithm can be found in Section \ref{Preliminaries}.}
	\label{fig:Intra-region-alt}
\end{figure}
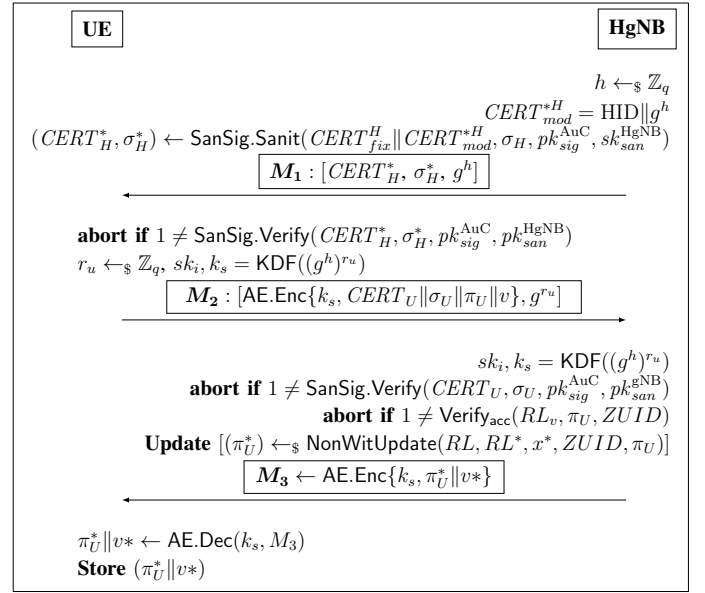
\fi

\begin{itemize}[leftmargin=*]

\item[]\textbf{Step $\mathbf{B_1}$:} HgNB $\rightarrow$ UE: $\mathbf{M_{1}}$: [$\cert_{H}^{*},\sigma^{*}_{H},g^{h}$].\\
This step proceeds identically to \textbf{Step} $\mathbf{A_1}$ of the initial authentication phase. In this regard, when a new user enters into the coverage area of new $\HgNB$, the $\HgNB$ sanitises his/her certificate, and composes a message $\mathbf{M_{1}}$, then sending $\mathbf{M_{1}}$ to UE.

\item[]\textbf{Step $\mathbf{B_2}$:} UE $\rightarrow$ HgNB:\\ $\mathbf{M_{2}}$:$[ \AuthEnc.\Enc\{k_s,\cert_U ||\sigma_U||\pi_U||v\},g^{r_{u}}]$ .\\
Upon receiving the message $\mathbf{M_{1}}$, UE verifies the $\HgNB$ certificate $\cert_{H}$ and the DH public keyshare $g^h$, using  $\SanSig.\Verify(\cert_{H}^{*}, \sigma^{*}_{H})$. If successful, $\UE$ samples $r_{u}$ and computes session keys $sk_{i},k_s$.
Next UE composes a message $\mathbf{M_{2}}$ and encrypts it using $k_s$. The encrypted part of $M_{2}$ consist of $\cert_U$, $\sigma_U$, $\pi_U$ and $v$, which is the user's certificate, certificate signature, non-membership witness and the accumulator version of which $\pi_U$ was created from, respectively. Finally, UE sends $\mathbf{M_{2}}$ to HgNB.

    \item [] \textbf{Step $\mathbf{B_3}$}: HgNB $\rightarrow$ UE: $\mathbf{M_{3}}$: [$\AuthEnc.\Enc\{k_s,\pi_U^*||v^*\}$].
    Upon receiving the response message $\mathbf{M_{2}}$, HgNB generates the session keys $sk_{i},k_s$, to decrypt $\mathbf{M_{2}}$. Subsequently, HgNB verifies $\UE$'s certificate using $\SanSig.\Verify(\cert_{U}, \sigma_{UE})$. If successful, HgNB recovers the accumulator version $v$ and checks if $v_i = v_{RL}$, to check if RL has been updated. If not, HgNB checks if the UE has been revoked by calling $\Verifyacc(ZUID_i,..)$. Otherwise, if the revocation list has been updated, where $v_i \neq v_{RL}$, HgNB checks if $ZUID_{i}$ has been accumulated in the updated RL. If not, HgNB updates the non-membership witness $\pi^{*}_{U}$ (where $x^{*}$ is the new unrevoked UE). Finally, HgNB encrypts and sends $\mathbf{M_{3}}$ to UE, which they will maintain for future communications. Details of this protocol is depicted in Figure \ref{fig:Intra-region-alt}.
\end{itemize}

\subsection{Inter-region Handover}

\begin{figure}[t!]
	\centering
	\begin{adjustbox}{max width=0.5\textwidth, max height=1\textheight}
		\fbox{
	\begin{tikzpicture}[yscale=-0.55,>=latex]
    \tikzstyle{every node}=[font=\large]
	\edef\InitX{0}
	\edef\ArrowLeft{1}
	\edef\ArrowCenter{6}
	\edef\ArrowRight{11}
	\edef\RespX{12}
	\edef\Y{0}
	
	\node [rectangle,draw,inner sep=5pt,right] at (\InitX,\Y) {\textbf{UE}};
	\node [rectangle,draw,inner sep=5pt,left] at ((7,\Y) {\textbf{HgNB}};
	\node [rectangle,draw,inner sep=5pt,left] at (\RespX,\Y) {\textbf{gNB}};
	\NextLine[2]
	\AdversaryAction{$h \getsr \mathbb{Z}_q$}
	\NextLine
	\AdversaryAction{$\cert^{*H}_{mod}=\HID\|g^{h}$}
	\NextLine
	\AdversaryAction{$(\cert_{H}^{*},\sigma^{*}_{H}) \gets \SanSig.\Sanit (\cert_{H},  \cert^{*H}_{mod},\sigma_{H}, \pk_{sig}^{\AuC}, \sk_{san}^{\HgNB} )$}
	\NextLine[1.5]
	\AdversaryToClient {\framebox[1.1\width]{$\boldsymbol{M_1}:[\cert_{H}^{*}$, $\sigma^{*}_{H}$, $g^h]$}}{}
	\NextLine
	\ClientAction{\textbf{abort if} $1 \neq \SanSig.\Verify(\cert_{H}^{*}, \sigma^{*}_{H}, \pk_{sig}^{\AuC}, \pk_{san}^{\HgNB} )$}
	\NextLine
	\ClientAction{$r_{u} \getsr \mathbb{Z}_q$, $sk_{i},k_s=\KDF((g^h)^{r_{u}})$}
	\NextLine[1.5]
	\ClientToAdversary{\framebox[1.1\width]{$\boldsymbol{M_2}:[\AuthEnc.\Enc\{k_s,\cert_{U}\|\sigma_U\| \pi_U\|v\},g^{r_u}]$}}{}
	\NextLine
	\AdversaryAction{$sk_{i},k_s=\KDF((g^h)^{r_{u}})$}
	\NextLine
	\AdversaryAction{$\cert_{U}\|\sigma_U\| \pi_U\|v \gets \AuthEnc.\Dec(k_s,M_{2})$}
	\NextLine[1.5]
	\AdversaryToServer{\framebox[1.1\width]{$\boldsymbol{M_3}:[\cert_{U}\|\sigma_U\| \pi_U\|v]$}}{}
	\NextLine
	\ServerAction{\textbf{abort if} $1 \neq \SanSig.\Verify(\cert_{U}, \sigma{U}, \pk_{sig}^{\AuC}, \pk_{san}^{\gNB} )$}
	\NextLine
	\ServerAction{\textbf{abort if} $1 \neq \mathit \Verifyacc(RL_v,\pi_U,ZUID )$}
	\NextLine
	\ServerAction{\textbf{Update} $ [(\pi^{*}_{U}) \getsr \Nwitupdate(RL,RL^{*},x^{*},ZUID, \pi_U)]$}
	\NextLine
	\ServerAction{$\RUID \getsr \mathcal{R}$}
	\NextLine
	\ServerAction{$\cert^{U^*}_{mod}=\RUID^*\|RID^*$}
	\NextLine
	\ServerAction{$(\cert_{U}^{*},\sigma^{*}_{U}) \gets \SanSig.\Sanit (\cert^{*U}_{fix},  \cert^{U^*}_{mod},\sigma_{U}, \pk_{sig}^{\AuC}, \sk_{san}^{\gNB} )$}
	\NextLine[1.5]
	\ServerToAdversary{\framebox[1.1\width]{$\boldsymbol{M_{4}:}[\sigma_{U^*}\|\cert_{U^*}\|\pi_{U^*} \|v^*]$}}{}
	\NextLine
	\AdversaryToClient{\framebox[1.1\width]{$\boldsymbol{M_{5}}\gets\AuthEnc.\Enc\{k_s,M_{4}\}$}}{}
	\NextLine
	\ClientAction{$\sigma^*_{U}\|\cert^*_{U}\|\pi_{U^*} \|v^*] \gets \AuthEnc.\Dec(k_e,M_{5}))$}
	\NextLine
	\ClientAction{\textbf{abort if} $\SanSig.\Verify(\sigma^*_{U},\cert^*_{U},\pk_{sig}^{\AuC},\pk_{san}^{\gNB})\neq1$}
	\NextLine
	\ClientAction{\textbf{Store} $(\pi^{*}_{U}\|v^*)$}
	\end{tikzpicture}
}
	\end{adjustbox}	\caption{The Inter-region Handover Phase of our proposed 5G Secure Handover Scheme. The descriptions of each algorithm can be found in Section \ref{Preliminaries}.}
	\label{fig:Inter-region-alt}
\end{figure}
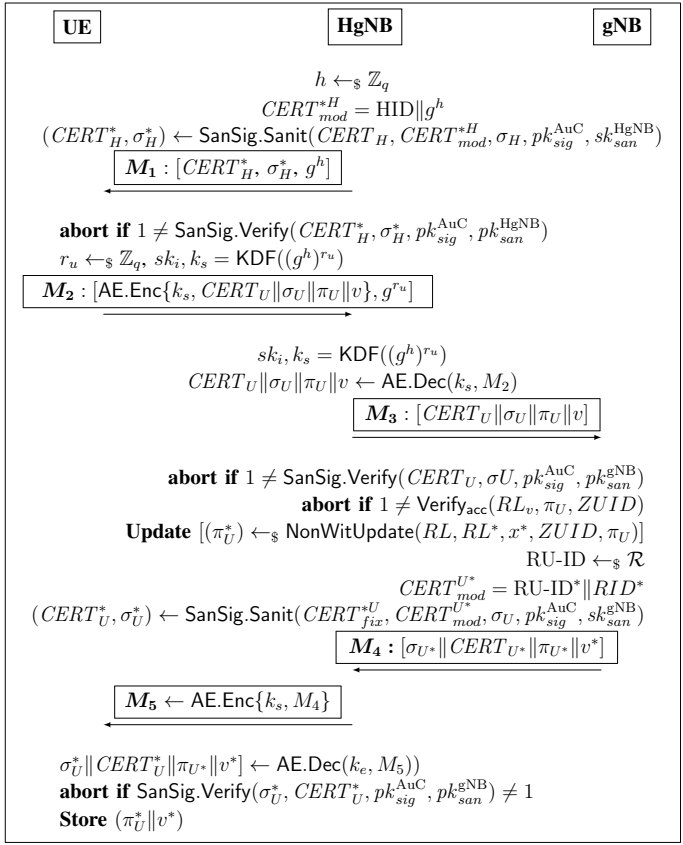

When a user moves to a different region, then they need to execute the inter-region handover phase of our proposed scheme, which is described below, and illustrated in Figure \ref{fig:Inter-region-alt}:

    \begin{itemize}[leftmargin=*]
    
    \item[]\textbf{Step $\mathbf{C_1}$:} HgNB $\rightarrow$ UE: $\mathbf{M_1}$: [$\cert_{H}^{*},\sigma^{*}_{H},g^{h}$].\\
    This step proceeds identically to \textbf{Step} $\mathbf{A_1}$ of the initial authentication phase. In this regard, for the new users entering the coverage area of $\HgNB$, the $\HgNB$ sanitises his/her certificate, and composes a message $\mathbf{M_{1}}$, then sending $\mathbf{M_{1}}$ to UE.
    
    \item[]\textbf{Step $\mathbf{C_2}$:} UE $\rightarrow$ HgNB:\\  $\mathbf{M_2}$:$[\AuthEnc.\Enc\{k_s,\cert_U||\sigma_U||\pi_U\||v\},g^{r_{u}}]$.\\
    This step proceeds identically to \textbf{Step} $\mathbf{B_2}$ of the intra-region HO phase. In this regard, $\UE$ verifies $\mathbf{M_1}$, compute a session key. $\UE$ then composes $\mathbf{M_2}$, encrypt it using the session key then send it to $\HgNB$. 
    
    \item []\textbf{Step $\mathbf{C_3}$}: HgNB $\rightarrow$ gNB: $\mathbf{M_3}$: [$\cert_U ||\sigma_U||\pi_U||v$].
    Upon the arrival of $\mathbf{M_2}$, HgNB generates the session keys $sk_{i},k_s$ to decrypt $\mathbf{M_2}$, then forwards the (decrypted) message to gNB.
    
    \item []\textbf{Step $\mathbf{C_4}$}: gNB $\rightarrow$ HgNB: $\mathbf{M_4}$: $[\sigma^*_{U}\|CERT^*_{U}\|\pi^*_{U}\|v*]$.
    \\After receiving the response message $\mathbf{M_3}$, gNB verifies the user's certificate using the $\SanSig$ verification algorithm, i.e.  $\SanSig.\Verify(\cert_{U},\sigma_U)$. 
    If successful, HgNB retrieves the accumulator version $v$ and checks if $v_i = v_{RL}$, to see if RL has been updated. If not, HgNB checks if the UE has been revoked by using $\Verifyacc(ZUID_i)$. Otherwise, if the revocation list has been updated (and $v_i \neq v_{RL}$) HgNB checks whether $ZUID_{i}$ is added in the later version of the RL. If not, HgNB updates the non-membership witness $\pi^{*}_{U} \gets \Nwitupdate(.)$ (where $x^{*}$ is the new non-revoked UE). Subsequently gNB updates the region-user identifier $RU-ID^*_i$, updates the ``modifiable'' region-specific part of the UE certificate $cert_{mod}^{*U}$, and updates the user certificate accordingly, where $\cert^*_{U}= cert_{mod}^{*U}||cert_{fix}^{U}$. After, gNB sanitises UE $\cert_{U} \gets \SanSig.\Sanit(.)$.  Finally, gNB composes $\mathbf{M_4}$, sending $\mathbf{M_4}$ to HgNB.

    \item []\textbf{Step $\mathbf{C_5}$}: HgNB $\rightarrow$ UE: $\mathbf{M_5}$: $[ \AuthEnc.\Enc\{k_s,M_{4}\}]$.
    Upon receiving the message $\mathbf{M_4}$, HgNB encrypts it  using the session key $k_s$, to generate $\mathbf{M_5}$, sending $\mathbf{M_5}$ to the UE.
 
    \item []\textbf{Step $\mathbf{C_6}$}: 
    Upon receiving the encrypted message $\mathbf{M_5}$, the UE recovers $(\sigma^*_{U},CERT^*_{U},\pi^*_{U},v*)$, and verifies the certificate signature, using $\SanSig$  verification algorithm i.e. $\SanSig.\Verify(CERT^*_{U},\sigma^*_{U})$. If verification fails, $\UE$ terminates the execution of the protocol. Otherwise, the user updates their certificate and RU-ID. Details of this protocol is depicted in Figure 4.
\end{itemize}

\section{Security Analysis}
\label{Security Analysis}
This section provides a formal proof that our protocols achieve mutual authentication, key indistinguishability and unlinkability. Note that each of our proofs proceeds as a series of game-hops, where we incrementally change the experiment and demonstrate at the end that the adversary cannot win (nor detect the changes) with non-negligible probability. We begin by analysing the $\MA$-security of each of our protocols in turn.

\subsection{Mutual Authentication security}
\label{sec:MA-game}
Here we discuss the mutual authentication security of our protocols. Due to space constraints, we only discuss the details of the proof of $\MA$-security for the Initial Authentication scheme since it's the most technically interesting. The full version of each proof is available in Appendix C of Supplementary Material. 
\begin{theorem}{\textbf{$\MA$-security of Initial Authentication}}.  Initial Authentication depicted in Figure \ref{fig:init-auth-alt} is $\MA$-secure under the cleanness predicate defined in [Appendix-B, Definition5]\footnote{The full version of the security analysis and the security framework is available in the Supplementary Material.}. For any PPT algorithm $\adv$ against the $\MA$ experiment, $\Adv{\MA,\cleanpredicate}{\Prot,\numParties,\numSessions,\adversary}(\secpar)$ is negligible assuming the EUFCMA security of $\SanSig$, Auth security of $\AuthEnc$, the KDF security of $\KDF$ and the $\DDH$ assumption.
\end{theorem}

\begin{pro} First, we recall that in order to win the $\MA$-security experiment, that $\adv$ cannot issue a $\Corrupt(i)$ query before a session $\testsess$ accepts such that $\chal$ terminates the game and outputs $1$, nor can it issue a $\StateReveal(i,s)$, nor a $\StateReveal(\AuC, s)$ query (where $\session_{\AuC}^{s'}$ received messages from $\testsess$). 

\iffullversion
We divide the proof into two cases: the first where the UE accepts messages $\mathbf{M_1}$ and $\mathbf{M_4}$ without an honest matching $\AuC$ partner. We denote $\adversary$'s advantage in \textbf{Case 1} as $\Adv{\MA,\cleanpredicate,\mathsf{C1}}{\Prot,\numParties,\numSessions,\adversary}(\secpar)$. The second case is when the AuC accepts messages $\mathbf{M_3}$ and $\mathbf{ACK}$ without an honest matching $\UE$ partner. We denote $\adversary$'s advantage in \textbf{Case 1} as $\Adv{\MA,\cleanpredicate,\mathsf{C2}}{\Prot,\numParties,\numSessions,\adversary}(\secpar)$. It is clear that \[\Adv{\MA,\cleanpredicate}{\Prot,\numParties,\numSessions,\adversary}(\secpar)\leq \Adv{\MA,\cleanpredicate,\mathsf{C1}}{\Prot,\numParties,\numSessions,\adversary}(\secpar) + \Adv{\MA,\cleanpredicate,\mathsf{C2}}{\Prot,\numParties,\numSessions,\adversary}(\secpar).\]\\
\textbf{Case 1: $\UE$ accepts messages $\mathbf{M_1}$ and $\mathbf{M_4}$ without an honest matching $\AuC$ partner.}

     \textbf{Game $A_{1.0}$}: This is the original mutual authentication experiment described in [Appendix B-A]\footnotemark[1]: \[\Adv{\MA,\cleanpredicate,\mathsf{C1}}{\Prot,\numParties,\numSessions,\adversary}(\secpar) \leq \Adv{}{G_{A_{1.0}}}\]

      \textbf{Game $A_{1.1}$}: Here we introduce an abort event, where the challenger aborts if $\adv$ produces a valid signature $\sigma$ that verifies under  $\pk_{sig}^{AuC}$ and $\pk_{san}^{HgNB}$. At the beginning of the experiment, we initialise a $\SanSig$ challenger $\challenger$, that outputs $\pk_{sig}^{\challenger}$ and  $\pk_{san}^{\challenger}$, which we embed into the AuC and HgNB respectively. Then any time AuC or HgNB needs to generate a signature, we query $\SanSig.\Sign$ to sign a message $m$. Now, we trigger the abort event that occurs whenever $\adv$ produces a valid signature. Thus, the probability that $\adv$ triggers the abort event is bounded by the EUFCMA security of $\SanSig$. Due to the page limit, the security of $\SanSig$ has been included in the [Appendix A, Definition 4] of the supplementary material : \[\Adv{}{G_{A_{1.0}}} \leq \Adv{}{G_{A_{1.1}}}+\Adv{\mathrm{EUFCMA}}{\SanSig}.\]
     
      \textbf{Game $A_{1.2}$}: In this game we guess the first session $\pi^s_i$ to accept without a matching partner, such that $\pi_i.role= UE$. Since there are at most $\numParties$ parties running $\numSessions$ sessions, the probability of session $\pi^s_i$ accepts without a matching partner is: \[\Adv{}{G_{A_{1.1}}} \leq \numParties \cdot \numSessions \cdot \Adv{}{G_{A_{1.2}}}.\]
     
     \textbf{Game $A_{1.3}$}: Here we introduce another abort event. That is triggers if $\adv$ sends a Diffie-Hellman public keyshare  $g^h$ to the session $\pi^s_i$, i.e. session $\pi^s_i$ receives $g^h$ that is not from an honest HgNB. Since this requires a signature over $g^h$, and by \textbf{Game} $A_{1.1}$ we abort if $\adv$ generates a valid signature over any message $m$, it follows that: \[\Adv{}{G_{A_{1.2}}} \leq \Adv{}{G_{A_{1.3}}}.\]
     
     \textbf{Game $A_{1.4}$}: In this game, we replace $g^h, g^{r_u}$ and $g^{hr_u}$ computed honestly in the protocol execution with $g^a, g^b, g^c$ respectively, from a DDH challenger. By the definition of Decisional Diffie-Hellman, $a,b,c$ are sampled uniformly at random from $\Zq$, and independent of the protocol execution. Thus any $\adversary$ that can distinguish \textbf{Game} $A_{1.3}$ from \textbf{Game} $A_{1.4}$ can break the DDH assumption [Appendix-A,Definition 1]\footnotemark[1]. Thus it follows that:  \[\Adv{}{G_{A_{1.3}}} \leq \Adv{}{G_{A_{1.4}}}+ \Adv{G,g,q}{\DDH}.\]
     
     \textbf{Game $A_{1.5}$}: In this game we replace the session and encryption keys $sk_i, k_s$ with uniformly random values $\hat{sk_i}, \hat{k_s}$ by interacting with a $\KDF$ challenger. Since $sk_i, k_s \gets \KDF(g^c)$ and by \textbf{Game $A_{1.4}$} $g^c$ is already uniformly random and independent, this change is sound. Any $\adversary$ that can distinguish \textbf{Game} $A_{1.4}$ from \textbf{Game} $A_{1.5}$ can be used to break $\KDF$ security defined in [Appendix A-Definition 1] on the supplementary material. Thus:  \[\Adv{}{G_{A_{1.4}}} \leq \Adv{}{G_{A_{1.5}}}+ \Adv{\mathrm{KDF}}{\KDF}.\]

     \textbf{Game $A_{1.6}$}: In this game, we introduce an abort event that occurs if $\testsess$ decrypts a valid ciphertext keyed by $\hat{sk_i}$, but the ciphertext was not produced by an honest AuC session. Specifically, we initialise an $\AuthEnc$ challenger that is queried whenever the challenger needs to encrypt or decrypt with $\hat{sk_i}$. The abort event only triggers if $\adv$ can produce a valid ciphertext, and we can submit the valid ciphertext to the $\AuthEnc$ challenger to break the security of the security of the AE scheme. By \textbf{Game $A_{1.5}$} $\hat{sk_i}$ is already uniformly random and independent and this replacement is sound. Any $\adversary$ that can trigger the abort event can break the Auth security of the $\AuthEnc$ scheme [Appendix A-Definition 2]\footnotemark[1].  This implies: \[\Adv{}{G_{A_{1.5}}} \leq \Adv{}{G_{A_{1.6}}}+ \Adv{\mathrm{Auth}}{\AuthEnc}.\]
     
     \textbf{Game $A_{1.7}$}: In this game, the session $\pi_i$ will only accept $\mathbf{M_1}$ from HgNB and $\mathbf{M_4}$ from AuC if they are honest partners. $\adv$ cannot produce a valid ciphertext by \textbf{Game} $A_{1.6}$, and $\adv$ cannot produce a valid signature by \textbf{Game} $A_{1.1}$. Thus the advantage of $\adv$ in winning the $\MA$-security experiment is negligible. 
    \[\Adv{}{G_{A_{1.7}}}=0.
    \]
    
\fi

\iffullversion
\textbf{Case 2: $\AuC$ accepts messages $\mathbf{M_3}$ and $\mathbf{ACK}$ without an honest matching $\UE$ partner.}
  In this case, we assume that the first session to accept without a matching partner is owned by AuC. 

     
     \textbf{Game $A_{2.0}$}: This is the original mutual authentication game described in [Appendix B-A]\footnotemark[1]: \[\Adv{\MA,\cleanpredicate,\mathsf{C2}}{\Prot,\numParties,\numSessions,\adversary}(\secpar) \leq \Adv{}{G_{A_{2.0}}}.\]
     
     \textbf{Game $A_{2.1}$} : In this game, we guess the index $i$ of the first $\AuC$ session that accepts without a matching partner such that their partner is owned by $\UE_i$, i.e. $\session_{\AuC}^{s}.\pid = \UE_i$, introducing a factor of $\numParties$ in $\adv$'s advantage: \[\Adv{}{G_{A_{2.0}}} \leq \numParties \cdot \Adv{}{G_{A_{2.1}}}.\]

     \textbf{Game $A_{2.2}$}: As per the definition of \textbf{Case 2}, $\adv$ cannot issue a $\Corrupt(i)$ query before the $\AuC$ session accepts without a matching partner. In this game, we introduce an abort event that triggers if the first $\session_{\AuC}^{t}$ to accept without a matching partner accepts a ciphertext $\mathbf{M_{3}}$ that was not output from a matching partner session $\testsess$. Specifically, we initialise an Auth challenger that is queried whenever $\challenger$ needs to encrypt or decrypt with $k_i$. The abort event only triggers if $\adv$ can produce a valid ciphertext, and we can submit the valid ciphertext to the Auth challenger to break the Auth security of the $\AuthEnc$ scheme. Since $k_i$ is uniformly random and cannot be leaked to $\adv$, this replacement is sound. Thus, any $\adversary$ that triggers this abort event can be used to break the Auth security of $\AuthEnc$ [Appendix A-Definition 2]\footnotemark[1], thus: \[\Adv{}{G_{A_{2.1}}} \leq \Adv{}{G_{A_{2.2}}}+ \Adv{\mathrm{Auth}}{\AuthEnc}.\]
     
     \textbf{Game $A_{2.3}$}: Here we introduce a similar abort event that triggers if $\session_{\AuC}^{t}$ accepts a ciphertext \textbf{ACK} that was not output from a matching partner session $\testsess$. The changes introduced to \textbf{Game} $A_{2.3}$ follow from \textbf{Game} $A_{2.2}$, and thus introduces no new advantage for $\adversary$. Thus: \[\Adv{}{G_{A_{2.2}}} \leq \Adv{}{G_{A_{2.3}}}.\]
 
     \textbf{Game $A_{2.4}$}: In this game, the $\session_{\AuC}^{t}$ only accepts $\mathbf{M_{3}}$ and $\mathbf{ACK}$ from an honest matching partner. Thus, summing the probabilities we find that the $\adv$ has negligible advantage in winning the $\MA$-security experiment.: 
        \[\Adv{}{G_{A_{2.4}}}=0
        \]
\fi

\end{pro}

The analysis of the $\MA$-security of the Intra- and Inter-region Handovers proceed very similarly, and due to space constraints we merely give the theorem statement in the main body, and point to the supplementary material for more details, in [Appendix C].
\begin{theorem}{\textbf{$\MA$-security of Intra-region and Inter-region Handover}}. The Intra-region and Inter-region Handover protocols described in Section \ref{Proposed scheme} are $\MA$-secure under the cleanness predicate defined in [Appendix B- Definition 5]\footnote[1]{The full version of the security analysis and the security framework is available in the Supplementary Material.}. For any PPT algorithm $\adv$ against the $\MA$ experiment, $\Adv{\MA,\cleanpredicate}{\Prot,\numParties,\numSessions,\adversary}(\secpar)$ is negligible assuming the EUFCMA security of $\SanSig$, Auth security of $\AuthEnc$, the KDF security of $\KDF$ and the $\DDH$ assumption.
\end{theorem}

\subsection{Unlinkability}
\label{sec:unlink-game}
Here we discuss the unlinkability security of our protocols. In these experiments, the $\adv$ can issue a $\Test(i,j)$ query, which initialises a new session $\pi_b$ owned by either party $i$, or party $j$, based on a bit $b$ sampled by the challenger. Thus, we consider a strong notion of anonymity where $\adv$ can win simply by linking the ``test'' session to another protocol execution where it knows the identity of the UE. Thus, we capture user anonymity, and unlinkability. Due to space constraints, we only discuss the details of the proof of $\Unlink$ security for the Inter-region Handover scheme, since its the most technically interesting. The analysis of the Initial Authentication and Intra-region protocols proceed identically. We begin by stating the results of our analysis for the Initial Authentication and Intra-region Handover scheme.  

\begin{theorem}{\textbf{$\Unlink$-security of Initial Authentication and Intra-region Handover}}.  Initial Authentication and Intra-region Handover protocols described in Section \ref{Proposed scheme} are unlinkable under the cleanness predicate available in [AppendixB-C,-Definition 9] of the supplementary material. For any PPT algorithm $\adv$ against the $\Unlink$ experiment, $\Adv{\Unlink,\cleanpredicate}{\Prot,\numParties,\numSessions,\adversary}(\secpar)$ is negligible assuming the EUFCMA security of $\SanSig$, Auth security of $\AuthEnc$, the KDF security of $\KDF$ and the $\DDH$ assumption.
\end{theorem}

Next, we analyse the $\Unlink$-security of the Inter-region Handover scheme.

\begin{theorem}{\textbf{$\Unlink$-security of Inter-region Handover}}.  The Inter-region Handover scheme depicted in Figure \ref{fig:Inter-region-alt} is unlinkable under the cleanness predicate in [Appendix B-Definition 9]\footnotemark[1]. For any PPT algorithm $\adv$ against the $\Unlink$ experiment, $\Adv{\Unlink,\cleanpredicate}{\Prot,\numParties,\numSessions,\adversary}(\secpar)$ is negligible assuming the EUFCMA security of $\SanSig$, Auth security of $\AuthEnc$, the KDF security of $\KDF$ and the $\DDH$ assumption.
\end{theorem}

\begin{pro}  First, we recall that in order to win the $\Unlink$-security experiment, that $\adv$ cannot issue a $\Corrupt(i)$ query before a session $\testsess$ accepts such that $\chal$ terminate the game and outputs $1$, nor can it issue a $\StateReveal(i,s)$, nor a $\StateReveal(\HgNB, s)$ query (where $\session_{\HgNB}^{s'}$ received messages from $\testsess$). As before, we proceed via a sequence of games.

\iffullversion
     
    \textbf{Game $B_0$}: This is the original unlinkability experiment described in [Appendix A-C]\footnotemark[1]: \[\Adv{\Unlink,\cleanpredicate}{\Prot,\numParties,\numSessions,\adversary}(\secpar) \leq \Adv{}{G_{B_0}}.\]
    
     \textbf{Game $B_{1}$}: In this game we introduce an abort event, where the challenger aborts if $\adv$ produces a valid signature $\sigma$ that verifies under $\pk_{sig}^{AuC}$ and $\pk_{san}^{HgNB}$ or $\pk_{sig}^{AuC}$ and $\pk_{san}^{gNB}$. At the beginning of this game, we initialise a pair of $\SanSig$ challengers which output $\pk_{sig}^{\challenger}$ and $\pk_{san}^{\challenger}$, which we embed into AuC, HgNB and gNB. Then every time AuC, HgNB or gNB needs to generate a signature, we query $\SanSig.\Sign$ to sign a message $m$. Now, we trigger the abort event that occurs whenever $\adv$ produces a valid signature. Thus, the probability that $\adv$ triggers the abort event is bounded by the EUFCMA security of $\SanSig$, defined in [Appendix A-Definition 4]\footnotemark[1]: \[\Adv{}{G_{B_0}} \leq \Adv{}{G_{B_{1}}}+2\cdot\Adv{\mathrm{EUFCMA}}{\SanSig}.\]
     
     \textbf{Game $B_2$}: In this game, we guess the first session $\testsess$ to accept without a matching partner, such that $\testsess.role= \UE$. We also introduce another abort event that triggers if $\adv$ sends a Diffie-Hellman public keyshare  $g^h$ to the session $\testsess$, i.e. session $\testsess$ receives $g^h$ that is not from an honest HgNB. Since this requires a signature over $g^h$, and by \textbf{Game} $B_{1}$ we abort if $\adv$ generates a valid signature over any message $m$, this introduces no additional bound. Since there are $\numParties$ parties running at most $\numSessions$ sessions, this introduces the following bound: \[\Adv{}{G_{B_{1}}} \leq \numParties \cdot \numSessions \cdot \Adv{}{G_{2}}.\]

     \textbf{Game $B_{3}$}: In this game, we replace $g^h, g^{r_u}$ and $g^{hr_u}$ computed honestly in the protocol execution with $g^a, g^b, g^c$ respectively, from a $\DDH$ challenger. By the definition of Decisional Diffie-Hellman, $a,b,c$ are sampled uniformly at random from $\Zq$, and independent of the protocol execution. Thus any $\adversary$ that can distinguish \textbf{Game} $B_{2}$ from \textbf{Game} $B_{3}$ can break the DDH assumption [Appendix A-Definition 1]\footnotemark[1]. Thus it follows that:  \[\Adv{}{G_{B_{2}}} \leq \Adv{}{G_{B_3}}+ \Adv{G,g,q}{DDH}.\]
     
     \textbf{Game $B_{4}$}: In this game the challenger replaces the session and encryption keys $sk_i, k_s$ with uniformly random values $\hat{sk_i}, \hat{k_s}$ by interacting with a $\KDF$ challenger. Since $sk_i, k_s \gets \KDF(g^c)$ and by \textbf{Game $B_{3}$} $g^c$ is already uniformly random and independent, this change is sound. Any $\adversary$ that can distinguish \textbf{Game} $B_{3}$ from \textbf{Game} $B_{4}$ can be used to break $\KDF$ security [Appendix A-Definition 1]\footnotemark[3]). Thus:   \[\Adv{}{G_{B_{3}}} \leq \Adv{}{G_{B_{4}}}+ \Adv{\mathrm{KDF}}{\KDF}.\]

     \textbf{Game $B_{5}$}: In this game, we replace the computation of the ciphertext $c = \Enc(sk_i,CERT_U \|\sigma_U\|\pi_U\|v)$ with $\hat{c} = \Enc(\hat{sk_i}, rand)$, where $rand \getsr \bits{L}$ and $L=|CERT_U \\ \|\sigma_U\|\pi_U\|v|$, and the computation of the ciphertext $c^* = \Enc(sk_i,CERT_U^* \|\sigma_U^*\|\pi_U^*\|v^*)$ with $\hat{c^*} = \Enc(\hat{sk_i}, rand^*)$ where $rand^* \getsr \bits{L^*}$ and $L^*=|CERT_U^* \|\sigma_U^*\|\pi_U^*\|v^*|$. We do so by interacting with an encryption challenger whenever $\hat{sk_i}$ is used by the challenger to encrypt a message, and issuing either an $\Enc$ oracle call $(CERT_U \|\sigma_U\|\pi_U\|v,rand)$ or $(CERT_U^* \|\sigma_U^*\|\pi_U^*\|v^*,rand^*)$. Note that if the bit $b$ sampled by the challenger is $0$, then we are in \textbf{Game $B_{4}$}, and otherwise we are in \textbf{Game $B_{5}$}.  Since $\hat{sk_i}$ is (by \textbf{Game $B_{5}$}) uniformly random and independent, this change is sound. Any adversary $\adversary$ that can distinguish between \textbf{Game $B_{4}$} and \textbf{Game $B_{5}$} can be used to break the security of $\AuthEnc$, and thus:  \[\Adv{}{G_{B_{4}}} \leq \Adv{}{G_{B_{5}}}+ \Adv{\mathrm{Conf}}{\AuthEnc}\]
     
     \textbf{Game $B_{6}$}: In this game we highlight that the channel between the HgNB and the gNB is assumed to be secure, thus $\adv$ cannot compromise any underlying plaintext sent between HgNB and gNB, and all messages sent to and from $\session_b$ are uniformly random and independent of the bit $b$ sampled  by the challenger. Thus it follows that $\adv$ has no advantage in guessing the bit $b$, and summing the probabilities $\adv$ has negligible advantage in winning the $\Unlink$ game. Thus: 
    \[\Adv{}{G_{B_{6}}}=0.
    \]
     
\end{pro}
\fi

\begin{table}
    \centering
    \caption{Features comparison.}
    \label{tab:sec.Features}
    \begin{tabular}{|p{0.17\linewidth}||p{0.1\linewidth}|p{0.1\linewidth}|p{0.1\linewidth}|p{0.1\linewidth}|p{0.13\linewidth}|}
    \hline
        \textbf{Features} & \textbf{MA} &\textbf{SA} & \textbf{PFS} & \textbf{SRM} & \textbf{SIHO} \\
        \hline \hline
        5G \cite{3rd_generation_partnership_project_3gpp_security_2020} & YES & NO & NO & NO & NO
        \\
        \hline
        ReHand \cite{fan_rehand_2020} & YES & YES & NO & YES & NO
        \\
        \hline
        RUSH \cite{zhang_robust_2021} & YES & Partial & Partial & NO & YES
        \\
        \hline
        LSHA \cite{yan_lightweight_2021} & YES & Partial & Partial & NO & NO
        \\
        \hline
        Ours & YES & YES & YES & YES & YES
        \\
    \hline \hline
    \multicolumn {6}{|c|}{\textbf{MA}:Mutual Authentication, \textbf{SA}:Strong Anonymity,}\\
    \multicolumn {6}{|c|}{\textbf{PFS}: Perfect Forward Secrecy, \textbf{SRM}: Secure Revocation}\\
    \multicolumn {6}{|c|}{ Management, \textbf{SIHO}: Secure Inter-region HO supports}\\
    \hline
    
    \end{tabular}

\end{table}

\subsection{Key Indistinguishability}
\label{sec:Key Ind-game}
Here we discuss the key indistinguishability security of our scheme. In the $\KIND$ game, the goal of $\adv$ is to distinguish between either the real key generated by the test session $\testsess$, or a completely random key sampled uniformly at random from the same distributing, capturing a strong notion of key secrecy. Due to space constraints, we only discuss the details of the proof of $\KIND$ security for the Intra-region Handover, since its the most concise, and all other proofs proceed similarly. 
We begin by stating the results of our analysis for the Initial Authentication and Inter-region Handover scheme. 

\begin{theorem}{\textbf{$\KIND$-security of Initial Authentication and Inter-region Handover}}.  Initial Authentication and Inter-region Handover protocols described in Section \ref{Proposed scheme} achieves $\KIND$-security under the cleanness predicate in [Appendix B-Definition 7]\footnotemark[1]. For any PPT algorithm $\adv$ against the $\KIND$ experiment, $\Adv{\KIND,\cleanpredicate}{\Prot,\numParties,\numSessions,\adversary}(\secpar)$ is negligible assuming the EUFCMA security of $\SanSig$, Auth security of $\AuthEnc$, the KDF security of $\KDF$ and the $\DDH$ assumption.
\end{theorem}

Next, we analyse the $\KIND$-security of the Intra-region Handover scheme.

\begin{theorem}{\textbf{$\KIND$-security of Intra-region Handover}}.  The Intra-region Handover scheme described in Figure \ref{fig:Intra-region-alt} is $\KIND$-secure under the cleanness predicate in [Appendix B-Definition 7]\footnotemark[1]. For any PPT algorithm $\adv$ against the $\KIND$ experiment, $\Adv{\KIND,\cleanpredicate}{\Prot,\numParties,\numSessions,\adversary}(\secpar)$ is negligible assuming the EUFCMA security of $\SanSig$, Auth security of $\AuthEnc$, the KDF security of $\KDF$ and the $\DDH$ assumption.
\end{theorem}

\iffullversion

\begin{table*}
\centering
\caption{Time Costs of Cryptography Operations.}
        \label{tab:crypto}
\begin{tabular}{|l|l|l|c|c|c|c|c|c|c|c|}
\hline
\textbf{Notation} &    $T_P$    &    $T_{SM}$   &    $T_{MSM}$     &    $T_{E}$     &         $T_{MAC}$      &     $T_{H}$     &    $T_{PRG}$    &  $T_{AES}$       &  $T_{ECC}$ &    $T_{Mod}$   \\ \hline
$T_{UE}$ (ms)  & 13.199 & 0.926 & 1.150 & 1.205 & 0.103 & 0.167 & 0.294  & 0.804 & 12.130 & 0.008 \\ \hline
$T_{Sys}$ (ms)  & 7.479 & 0.235 & 0.294 & 0.340 & 0.071 & 0.089 & 0.1273 & 0.427 & 10.922 & 0.001 \\ 
    \hline \hline
    \multicolumn {11}{|c|}{$T_P$: pairing operation, $T_{SM}$: scalar multiplication, $T_{MSM}$: multi elliptic curve scalar multiplication,}\\
    \multicolumn {11}{|c|}{$T_{E}$: exponentiation operation, $T_{MAC}$: Massage authentication operations(Hmac-SHA256),}\\
    \multicolumn {11}{|c|}{$T_{H}$: Hash operations(SHA-256), $T_{PRG}$: Random number generators,$T_{AES}$:Symmetric encryption/decryption }\\
    \multicolumn {11}{|c|}{operations, $T_{ECC}$: Asymmetric encryption/decryption operations, $T_{Mod}$: Modular operations}\\
    \hline
\end{tabular}

\end{table*} 

\begin{table*}
\centering
    \caption{Performance comparison based on computational cost \\(Initial authentication).}
        \label{tab:Per. IA}
\begin{tabular}{|l|l|l|l|}
\hline
\textbf{Protocol}   & \textbf{Entity} & \textbf{Initial authentication} & \textbf{\makecell[l]{Total \\ time (ms)}}\\ \hline
\multirow{2}{*}{Conventional 5G-AKA\cite{3rd_generation_partnership_project_3gpp_security_2020}}     & $T_{UE}$ & $4{T_{PRG}} + 2{T_{MAC}} + {T_{AES}}$ &   $\approx$ 2.186    \\ \cline{2-4} 
                        &$T_{Sys}$&$3{T_{PRG}} + 1{T_{MAC}} + 2{T_H} + {T_{ECC}}$&  $\approx$ 11.553     \\ \hline
\multirow{2}{*}{ReHand\cite{fan_rehand_2020}} &$T_{UE}$&$2{T_{AES}} + 4{T_H} + {T_{PRG}}$&   $\approx$ 2.57    \\ \cline{2-4} 
                        &$T_{Sys}$&$3{T_{AES}} + 5{T_H} + {T_{PRG}}$&  $\approx$ 1.8533     \\ \hline
\multirow{2}{*}{RUSH\cite{zhang_robust_2021}}   &$T_{UE}$& $7{T_{PRG}} + 2{T_{MAC}} + {T_{ECC}} +3{T_H} + {T_E}+ 5{T_{Mod}} + 3{T_{SM}}$ &  $\approx$ 18.918     \\ \cline{2-4} 
                        &$T_{Sys}$& $4{T_{PRG}} + 1{T_{MAC}} + 3{T_H} + {T_{ ECC}} + 2{T_{SM}} + 3{T_{Mod}}$&   $\approx$ 12.2422    \\ \hline
\multirow{2}{*}{LSHA\cite{yan_lightweight_2021}}   &$T_{UE}$&$4{T_{PRG}} + 2{T_{MAC}} + {T_{AES}}$&  $\approx$ 2.188    \\ \cline{2-4} 
                        &$T_{Sys}$&$3{T_{PRG}} + 1{T_{MAC}} + 2{T_H} + {T_{ECC}}$&  $\approx$ 11.5529     \\ \hline
\multirow{2}{*}{OURs}   &$T_{UE}$&$3{T_{PRG}} + 5{T_{AES}} + {T_H} + {T_P}+ 7{T_E} + {T_{Mod}}$&  $\approx$ 26.631    \\ \cline{2-4} 
                        &$T_{Sys}$&$ 7{T_{PRG}} + 3{T_H} + 5{T_{AES}} + 10{T_E} + 2{T_{Mod}}$&   $\approx$ 6.693   \\ \hline
\end{tabular}

\end{table*}

\begin{table*}
\centering
\caption{Performance comparison based on computational cost \\(Intra-region Handover).}
        \label{tab:Per. Intra}
\begin{tabular}{|l|l|l|l|}
\hline
\textbf{Protocol}  & \textbf{Entity} & \textbf{Intra-region Handover} & \textbf{Total time (ms)} \\ \hline
\multirow{2}{*}{Conventional 5G-AKA \cite{3rd_generation_partnership_project_3gpp_security_2020}}     &$T_{UE}$&$4{T_{AES}} + {T_{PRG}}$&    $\approx$ 3.51   \\ \cline{2-4} 
                        &$T_{Sys}$&$4{T_{AES}} + {T_{PGR}}$&   $\approx$ 1.835    \\ \hline
\multirow{2}{*}{ReHand \cite{fan_rehand_2020}} &$T_{UE}$&${T_{PGR}} + 3{T_H} + {T_{AES}}$&  $\approx$ 1.599     \\ \cline{2-4} 
                        &$T_{Sys}$&${T_{PRG}} + 5{T_H} + 2{T_{AES}}$&   $\approx$ 1.426    \\ \hline
\multirow{2}{*}{RUSH \cite{zhang_robust_2021}}   &$T_{UE}$&$3{T_{PRG}} + {T_{SM}} + 5{T_H} + {T_E} + {T_{Mod}}$&   $\approx$ 2.856    \\ \cline{2-4} 
                        &$T_{Sys}$&$2{T_{PRG}} + {T_{SM}} + 6{T_H} + {T_E} + {T_{Mod}}$&   $\approx$ 1.365   \\ \hline
\multirow{2}{*}{LSHA \cite{yan_lightweight_2021}}   &$T_{UE}$&$2{T_{PRG}} + {T_{MAC}} + 4{T_{AES}}$&  $\approx$ 3.907    \\ \cline{2-4} 
                        &$T_{Sys}$&$7{T_{PRG}} + {T_{MAC}} + 6{T_{AES}} + 12{T_{Mod}}$&   $\approx$ 3.536    \\ \hline
\multirow{2}{*}{OURs}   &$T_{UE}$&$2{T_{PGR}} + 2{T_{AES}} + {T_h} + {T_E} + {T_{Mod}}$& $\approx$ 3.555     \\ \cline{2-4} 
                        &$T_{Sys}$&$5{T_{PGR}} + 2{T_{AES}} + {T_P} + {T_h} + 7{T_E} + {T_{Mod}}$& $\approx$ 11.42    \\ \hline
\end{tabular}

\end{table*}

\begin{table*}
\centering
\caption{Computational cost of the Inter-region Handover.}
        \label{tab:Per. Inter}
\begin{tabular}{|l|l|l|l|}
\hline
\textbf{Protocol} & \textbf{Entity} & \textbf{Inter-region Handover} & \textbf{Total time (ms)} \\ \hline
\multirow{2}{*}{OURs}   &$T_{UE}$&$ 2{T_{PGR}} + 2{T_{AES}} + {T_H} + {T_P} + 7{T_E} + {T_{Mod}}$&   $\approx$ 23.96   \\ \cline{2-4} 
                        &$T_{Sys}$&$ 6{T_{PGR}} + 2{T_{AES}} + 2{T_H}+  {T_P} + 15{T_E} + {T_{Mod}}$&  $\approx$  14.37   \\ \hline
\end{tabular}
    
\end{table*}

\begin{pro}
 We proceed via a sequence of games.

    \textbf{Game $C_0$}: This is the original $\KIND$ experiment described in [Appendix B-B]\footnotemark[1]: \[\Exp{\KIND,\cleanpredicate}{\Prot,\numParties,\numSessions,\adversary}(\secpar) \leq Adv_{G_{C_0}}.\]

    \textbf{Game $C_1$}: In this game, we introduce an abort event that triggers if $\chal$ guesses a session $\session_{i^*}^{s^*}$ and $\adv$ issues a $\Test$ query to a session $\testsess$ where $i\neq i^*$ and $s \neq s^*$. Since there are at most $(\numSessions \cdot \numParties)$ such sessions, this introduces the bound: \[\Adv{}{G_{C_0}} \leq \numSessions \cdot \numParties \cdot \Adv{}{C_{C_1}}.\]

    \textbf{Game $C_2$}: Here we introduce a new abort event, if the test session $\testsess$ accepts any message from a non-honest $\HgNB$ / $\UE$. This exactly matches the  $\MA$-experiment, where the $\adv$ attempts to inject messages from HgNB or UE. Thus, this game is bounded by the probability of $\adv$ breaking the $\MA$-security of the Intra-region HO phase, and thus:
    \[\Adv{}{G_{C_{1}}} \leq  \Adv{\MA,\cleanpredicate}{\Prot,\numParties,\numSessions,\adversary}(\secpar) + \Adv{}{G_{C_2}}\]
     
     \textbf{Game $C_3$}: In this game, we replace $g^h, g^{r_u}$ and $g^{hr_u}$ computed honestly in the protocol execution with $g^a, g^b, g^c$ respectively, from DDH challenger. By the definition of Decisional Diffie-Hellman, $a,b,c$ are sampled uniformly at random from $\Zq$, and independent of the protocol execution. Thus any $\adversary$ that can distinguish \textbf{Game} $C_{2}$ from \textbf{Game} $C_{3}$ can break the DDH assumption [Appendix A-Definition 1]\footnotemark[1]. Thus it follows that:  \[\Adv{}{G_{C_{2}}} \leq \Adv{}{G_{C_3}}+ \Adv{G,g,q}{DDH}.\]
     
    \textbf{Game $C_4$}: In this game the challenger replaces the encryption and session keys $sk_i, k_s$ with uniformly random values $\hat{sk_i}, \hat{k_s}$ by interacting with a $\KDF$ challenger. Since $sk_i, k_s \gets \KDF(g^c)$ and by \textbf{Game $C_{3}$} $g^c$ is already uniformly random and independent, this change is sound. Any $\adversary$ that can distinguish \textbf{Game} $C_{3}$ from \textbf{Game} $C_{4}$ can be used to break $\KDF$ security [Appendix A-Definition 1]\footnotemark[3]). Thus:   \[\Adv{}{G_{C_{3}}} \leq \Adv{}{G_{C_{4}}}+ \Adv{\mathrm{KDF}}{\KDF}.\]
    
    \textbf{Game $C_5$}:
    We highlight that as a result of these changes, the session key $\hat{sk_i}$ is now uniformly random and independent of the protocol execution regardless of the bit $b$ sampled by $\chal$, thus $\adv$ has no advantage in guessing the bit $b$:
    \[\Adv{}{G_{C_5}}=0.
    \]


\fi
\end{pro}

\section{Performance Evaluation and Comparison}
\label{sec: Performance}
The main objective of the proposed scheme is to provide the required security properties  (as discussed in Section \ref{subsec: Security Goals}) and to ensure a reasonable computational overhead. We compare the security of our scheme with state of the art protocols introduced in previous literature \cite{3rd_generation_partnership_project_3gpp_security_2020},\cite{fan_rehand_2020},\cite{zhang_robust_2021},\cite{yan_lightweight_2021}. Table \ref{tab:sec.Features} shows that all previously proposed schemes achieve mutual authentication but most \cite{3rd_generation_partnership_project_3gpp_security_2020},\cite{fan_rehand_2020},\cite{zhang_robust_2021},\cite{yan_lightweight_2021} fail to achieve \emph{all} security properties required of 5G.  For instance, 5G-AKA does not support strong anonymity and forward secrecy. Thus, schemes that use the original 5G-AKA protocol as their initial authentication phase can not support the strong anonymity and forward secrecy. RUSH \cite{zhang_robust_2021} and LSHA \cite{yan_lightweight_2021} protocols, for example, support PFS and anonymity in handover. Nevertheless, due to their dependency on the standard 5G-AKA, their overall schemes provide only partial support for the former security properties.
Next, from Table \ref{tab:sec.Features}, we see that only ReHand\cite{fan_rehand_2020} provides revocation management. Still, the performance of their revocation decreases when the number of users increases, which is required for the scalability that 5G schemes must achieve. While LUSH\cite{zhang_robust_2021} achieves inter-region HO, their proposed solution requires blockchain technologies, and the security and the performance of the blockchain in the authentication and handover schemes have been overlooked. In contrast, our proposed scheme can achieve a secure inter-region HO without the use of blockchain.

Next, we evaluate and compare the performance of the proposed scheme with the existing related works. We assume that all the aggregated network entities (AuC, HgNB, gNB) have higher computational capabilities than UE. Now, we conduct simulations of the cryptographic operations used by various schemes on a Dell Inspiron machine with i7 core, 2.30GHz CPU and 16.0 GB RAM (operating as the aggregated network entities as per the scheme).
To simulate the UE/smartphone, we use the Android studio emulator (Nexus 6) that runs Android API Tiramisu and is equipped with 1.5GB RAM with a six-core 3.0GHz processor. To implement the required cryptographic operations of our proposed scheme and the related works we used the java pairing-based cryptography (JPBC) \cite{de_caro_jpbc_2011} and Java Cryptography Extension (JCE) \cite{technology_network_java_nodate}. Table \ref{tab:crypto} shows the computation cost of the underlying cryptographic primitives, and use this to measure the overall computational cost of the protocols.

Table \ref{tab:Per. IA} presents the computational cost of the initial authentication protocols, which shows that the conventional 5G-AKA protocol is cheapest in terms of computational cost. However, the conventional 5G-AKA protocol does not support strong anonymity and forward secrecy properties. On the contrary, our proposed protocol achieves all security properties (demonstrated in Section \ref{Security Analysis} and Appendix-C of the supplementary material). Our proposed scheme introduces computational overhead compared to existing solutions, as shown in Table \ref{tab:Per. IA} and Table \ref{tab:Per. Intra}, required to achieve stronger security properties.  Furthermore, our schemes are more flexible, supporting more secure handover settings in the 5G-based mobile communication environments. Finally, Table \ref{tab:Per. Inter} presents the computational cost of inter-region HO of our proposed protocol only. Although RUSH supports the inter-region HO using blockchain, as previously discussed, the computational impact of using blockchain technologies in RUSH has been overlooked. For the Inter-region HO, our protocol requires approximately 23.96 ms and 14.37 ms on UE and system components (HgNB, gNB), respectively. Since we are the \emph{first}, up to our knowledge, that provides a secure inter-region HO solution without the assistance of a third party (AuC) or the intervention of external technology (blockchain), we consider this to be an acceptable performance for our setting. Additionally, our scheme targets the SCN scenario, which reduces the geographical distance between the users and BSs. Hence, latency generated by the geographical distance is reduced in the SCN scenario in 5G, which plays in favour of our scheme's latency. In order to further reduce the online computational cost, the network components (AuC, gNB and HgNB) can perform \emph{Batch Verification} for verifying users' certificates, where the batch verification can be defined as a method for verifying a large number of digital signatures simultaneously to accelerate signature verification process. The computational overhead of both signing and verifying of sanitisable signatures requires several pairing operations, which requires significant processing time. To resolve this issue, in our scheme, both the signers and verifiers (i.e., network components and UE) can pre-compute the majority of the required pairing operations for signing and verifying sanitisable signatures. In this way, our proposed scheme can reduce the number of pairing operations during the execution of the protocols, and that optimises the computational cost.



\section{Conclusion}
\label{Conclusion}
This paper highlights a need for a secure handover scheme that supports seamless mobility in 5G, specifically for a setting that supports SCNs. We evaluate the limitations of the existing solutions, which are insufficient for realising 5G stringent requirements of secure, privacy-preserving and reliable handover authentication mechanisms. This paper introduces an asymmetric-key-based authenticated key exchange and handover protocol that preserves user privacy and network security while providing a seamless region-based handover mechanism and an effective membership revocation management. We prove that all proposed protocols achieve strong mutual authentication, key indistinguishability and strong user anonymity, through the use of the underlying secure cryptographic functions, such as sanitisable signatures, ephemeral Diffie-Hellman key exchange, key derivation functions and authenticated encryption. The proposed 'privacy-aware secure region-based handover protocol for small-cell networks in 5G-enabled mobile communication' achieved the required security properties for roaming users in SCN 5G networks. Finally, we evaluated the performance of our proposed schemes, and compared them with existing solutions for 5G, and demonstrated that our schemes compare well. As for future work, we will be focusing on addressing the heterogeneity of the 5G networks (HetNets) with SCN by designing a cross-layer-based authentication protocol.

\bibliography{mybibliography}


\end{document}